\documentclass[a4paper, 11pt, oneside, halfparskip]{scrartcl}
\usepackage[utf8]{inputenc}
\usepackage{serif-report}
\usepackage{mathpazo}
\usepackage{color, colortbl}
\usepackage{booktabs}
\usepackage{longtable}

\usepackage{iftex}
\ifPDFTeX
  \usepackage[T1]{fontenc}
  \usepackage[utf8]{inputenc}
  \usepackage{textcomp} 
\else 
  \usepackage{unicode-math} 
  \defaultfontfeatures{Scale=MatchLowercase}
  \defaultfontfeatures[\rmfamily]{Ligatures=TeX,Scale=1}
\fi
\usepackage{lmodern}
\ifPDFTeX\else
\fi
\IfFileExists{upquote.sty}{\usepackage{upquote}}{}
\IfFileExists{microtype.sty}{
  \usepackage[]{microtype}
  \UseMicrotypeSet[protrusion]{basicmath} 
}{}
\makeatletter
\@ifundefined{KOMAClassName}{
  \IfFileExists{parskip.sty}{%
    \usepackage{parskip}
  }{
    \setlength{\parindent}{0pt}
    \setlength{\parskip}{6pt plus 2pt minus 1pt}}
}{
  \KOMAoptions{parskip=half}}
\makeatother
\usepackage{color}
\usepackage{fancyvrb}

\DefineVerbatimEnvironment{Highlighting}{Verbatim}{commandchars=\\\{\}}
\newenvironment{Shaded}{}{}

\newcommand{\AttributeTok}[1]{\textcolor[rgb]{0.49,0.56,0.16}{#1}}

\newcommand{\BuiltInTok}[1]{\textcolor[rgb]{0.00,0.50,0.00}{#1}}

\newcommand{\ControlFlowTok}[1]{\textcolor[rgb]{0.00,0.44,0.13}{\textbf{#1}}}
\newcommand{\DataTypeTok}[1]{\textcolor[rgb]{0.56,0.13,0.00}{#1}}

\newcommand{\ExtensionTok}[1]{#1}

\newcommand{\FunctionTok}[1]{\textcolor[rgb]{0.02,0.16,0.49}{#1}}

\newcommand{\KeywordTok}[1]{\textcolor[rgb]{0.00,0.44,0.13}{\textbf{#1}}}
\newcommand{\NormalTok}[1]{#1}

\newcommand{\PreprocessorTok}[1]{\textcolor[rgb]{0.74,0.48,0.00}{#1}}

\newcommand{\VariableTok}[1]{\textcolor[rgb]{0.10,0.09,0.49}{#1}}

\usepackage{longtable,booktabs,array}
\usepackage{calc} 
\usepackage{etoolbox}
\makeatletter
\patchcmd\longtable{\par}{\if@noskipsec\mbox{}\fi\par}{}{}
\makeatother
\IfFileExists{footnotehyper.sty}{\usepackage{footnotehyper}}{\usepackage{footnote}}
\makesavenoteenv{longtable}
\providecommand{\tightlist}{%
  \setlength{\itemsep}{0pt}\setlength{\parskip}{0pt}}
\usepackage{bookmark}
\IfFileExists{xurl.sty}{\usepackage{xurl}}{} 

\title{AutoAppendix: Towards one-click {Reproduction} of {Computational} {Artifacts}}
\author{Klaus Kra\ss nitzer\\TU Wien\\
\ttt{krassnitzer@par.tuwien.ac.at}}
\date{August 30, 2024}

\hypersetup{
    colorlinks=true,
    linktoc=all,
    pdftitle={AutoAppendix: Towards one-click Reproduction of Computational Artifacts},
    pdfauthor={Klaus Krassnitzer},
    pdfkeywords={reproducibility, computational artifacts, supercomputing conference, SC24, Chameleon Cloud}
}

\renewcommand{\authorshort}{Klaus Kraßnitzer}
\definecolor{Gray}{gray}{0.9}

\begin{document}
\maketitle
\begin{abstract}
\noindent \textbf{Abstract}. This report summarizes the findings of the AutoAppendix project, conducted
during the UCSC OSPO Summer of Reproducibility 2024. The project involved a
evaluation of reproducibility artifacts submitted to SC24,
focusing on their deployability and robustness on the Chameleon Cloud platform.
This technical report aims to inform and
support the reproducibility community by sharing observed challenges, patterns,
and best practices. Furthermore, we share templates developed
for Chameleon Cloud's Jupyter interface that are intended
to assist future authors and reviewers in streamlining artifact
evaluation workflows.
\end{abstract}
\section{Introduction}
The AutoAppendix project is a strategic effort to enhance artifact evaluation
processes for the Supercomputing (SC) Conference Series. This initiative is
rooted in the recognition that high-performance computing (HPC) is a domain
where scientific rigor and reproducibility are paramount. The SC Conference
Series has been championing these values by promoting transparency and
replicability through its reproducibility initiative, in which independent
reviewers evaluate the reproducibility of computational artifacts submitted
alongside research papers.

Alongside paper submissions, authors can apply for reproducibility badges
awarded by the SC Reproducibility Committee. These badges serve as marks of
quality and reliability, assuring readers that, given the necessary hardware,
the key results of a paper can be reproduced by others. Chameleon Cloud \cite{keahey_lessons_2020} -- a
large-scale, reconfigurable testbed for cloud computing research -- has been
instrumental in supporting these reproducibility efforts. Its configurability
and accessibility enable a wide range of experiments, making Chameleon an ideal
platform for our reproduction of artifact workflows.

The primary aim of AutoAppendix is to conduct an exhaustive evaluation of the
Artifact Description/Artifact Evaluation (AD/AE) appendices submitted to the
Supercomputing Conference 2024 (SC24) \cite{noauthor_sc_2024}, particularly those utilizing Chameleon
Cloud. We explore the current capabilities and limitations of
these appendices in fostering reproducibility. By identifying where greater
automation could be introduced, the project seeks to streamline artifact
reproduction and reduce human error.

In this report, we document the project’s journey from conception to
culmination, detail the methodologies used to assess the appendices, and discuss
steps to improve the reproducibility workflow for authors and reviewers. We also
propose new tools for the Chameleon Cloud platform aimed at simplifying the
creation and evaluation of computational artifacts. A comprehensive
documentation of our artifact evaluations is provided in the Appendix.

\section{Selection of relevant submissions}
To evaluate potential improvements in artifact reproducibility, we selected a
subset of submissions from SC24. This section outlines our selection process and
the criteria for determining each submission’s relevance to the AutoAppendix
project. Because our goal was to enhance the reproducibility of computational
artifacts, we concentrated on single-node experiments that could be easily
replicated on Chameleon Cloud without specialized hardware or complex network
configurations.

In this report, we refer to each appendix by an anonymized ID (e.g., \texttt{apdx111}).
These IDs were assigned during the double-blind review process and are not publicly visible.
\footnote{We are happy to provide non-anonymized references or additional details on request.}

\subsection{Understanding SC24 Reproducibility Badges}

For SC24, authors can apply for reproducibility badges based on the ACM
Reproducibility Standard \cite{acm_artifact_2020, noauthor_adae_nodate}. These badges recognize and encourage practices that
improve the transparency and reproducibility of research results. There are
three main types of badges:
\begin{itemize}
\item \textbf{Artifacts Available}: Awarded when the artifacts (software, data,
etc.) underlying the paper are publicly available in an archival repository.
This ensures other researchers can access the materials needed to understand or
build upon the work.
\item \textbf{Artifacts Evaluated – Functional}: Indicates that the provided
artifacts are well-documented, complete, and have been verified as executable.
This goes beyond mere availability by confirming the artifacts function as
claimed in the research.
\item \textbf{Results Replicated}: The highest level of recognition, granted
when an independent person or team successfully replicates the paper’s main
results using the provided artifacts. This badge provides rigorous validation of
the findings, demonstrating their reliability under different conditions.
\end{itemize}

\subsection{Selection Process for Relevant Papers}

Our project focused on submissions that applied for the ``Results Replicated''
badge, as these represent the pinnacle of reproducibility and transparency at
SC24. Out of 45 submissions seeking this badge, we selected 18 that were most
relevant to our objectives.

Before finalizing our choices, we classified each candidate paper by its
relevance (the number in parentheses indicates the count in each category):
\begin{itemize}
\item \textbf{Immediately Relevant} (7): Artifacts directly aligned with the
project’s objectives and designed to work on Chameleon Cloud.
\item \textbf{Relevant} (11): Papers related to the project’s objectives but not
as directly aligned. They still offer valuable insights into reproducibility in
computational research.
\item \textbf{Potential} (4): Artifacts whose viability on Chameleon Cloud needs
further evaluation. These may be considered for inclusion pending additional
analysis of their relevance.
\item \textbf{Hardware Constrained} (11): Papers with experiments requiring
specific hardware configurations or resources not readily available on Chameleon
Cloud.
\item \textbf{Semi-Hardware Constrained} (3): Papers with experiments that could
be adapted to Chameleon Cloud, but requiring additional effort or modifications
to the setup.
\item \textbf{Not Relevant} (9): Papers not directly related to the project’s
objectives and requiring hardware unavailable on Chameleon Cloud (e.g., special
GPUs or supercomputers).
\end{itemize}

We ultimately narrowed our focus to submissions emphasizing single-node
experiments, particularly those easily replicable on Chameleon Cloud. With input
from the project mentor and based on the above classification, we finalized a
selection of 18 papers (``Immediately Relevant'' and ``Relevant'' categories).

The main objective in evaluating these selected artifacts—especially those
aiming for a ``Results Replicated'' badge—was to verify the robustness of their
software setups and dependencies, rather than to exactly reproduce all results.
As a byproduct, our evaluation results can help validate and cross-check the
official reviewers’ findings for these artifacts.

By emphasizing solid software setups and well-managed dependencies, we aim to
streamline the replication process and set a higher standard for future artifact
submissions. Ensuring that artifacts are well-documented and easily configurable
will help reviewers and other researchers engage with them more effectively,
fostering a stronger culture of reproducibility and rigorous validation. We
anticipate that this approach will lead to more reliable research outcomes and
greater trust in the findings from SC24 onward.

\section{Evaluation of Computational Artifacts}

We evaluated the selected appendices over approximately five weeks, from late
July 2024 to the end of August 2024. All evaluations were conducted on Chameleon
Cloud using a variety of hardware types (mostly running Ubuntu 22.04). Every
evaluation was documented in detail; the complete log of this process is
provided in the Appendix. Table \ref{tab:nodes} gives an overview of the
used hardware configurations and project types of the evaluated artifacts.

\begin{table}[h]
    \centering
    \caption{Nodes and project types for each artifact}
    \label{tab:nodes}
    \begin{tabular}{cccc}
        \toprule
        \textbf{ID} & \textbf{Node Type} & \textbf{Site} & \textbf{Project Type} \\
        \midrule
        111 & \texttt{skylake} & TACC & Simulator \\
        174 & \texttt{skylake} & TACC & Plotting \\
        190 & \texttt{skylake} & TACC & Plotting \\
        191 & \texttt{icelake\_r650} & TACC & OpenMP \\
        202 & \texttt{icelake\_r650} & TACC & Algorithm/Study \\
        227 & \texttt{v100} & UC & CUDA \\
        332 & \texttt{gtx\_6000} & UC & CUDA \\
        362 & \texttt{gtx\_6000} & UC & CUDA \\
        368 & \texttt{icelake\_r650} & TACC & Framework \\
        376 & \texttt{p100} & TACC & CUDA \\
        407 & \texttt{mi100} & TACC & OpenCL \\
        428 & \texttt{icelake\_r650} & TACC & Study \\
        466 & \texttt{v100} & UC & Study \\
        467 & \texttt{p100} & TACC & Framework \\
        482 & \texttt{icelake\_r650} & TACC & Algorithm/Study \\
        483 & \texttt{icelake\_r650} & TACC & Framework \\
        496 & \texttt{icelake\_r650} & TACC & Algorithm \\
        506 & \texttt{icelake\_r650} & TACC & Compiler \\
        \bottomrule
    \end{tabular}
\end{table}

Evaluating the computational artifacts was time-intensive and required careful
planning. Although we had essentially unlimited computation time for this
project, different appendices demanded different hardware resources. GPU
availability was a particular bottleneck (especially for NVIDIA A100 GPUs),
which made it challenging to schedule long-running experiments. In some cases,
multiple reservation attempts were needed to successfully run an experiment to
completion.

The process was further hampered by the fact that many authors did not test
their artifacts on a fresh system or failed to provide adequate documentation.
We encountered numerous issues, including hardware incompatibilities, failed
software installations, and missing experiment data. Even when an experiment ran
to completion, the expected results were often not clearly stated, making it
difficult to determine whether we had successfully replicated the paper’s
results.

These challenges raise the question of how the reproducibility process can be
improved. Notably, for each problematic artifact with unclear documentation or a
broken setup, we found at least one artifact that was well-documented and easy
to execute. We observed a trend toward containerization: several artifacts were
packaged as Docker, VirtualBox, or Singularity images. Most of these
containerized solutions worked out of the box, though some required additional
setup steps.

Another trend was to break the experiment into multiple steps with dedicated
scripts for each step. Often these steps were very granular (e.g., a separate
script for each dependency installation), but this approach made it easier for a
reviewer to understand the experiment’s components. This modularization also
paves the way for potential full automation of the artifact evaluation process,
which is the ultimate goal of reproducibility initiatives.

One appendix, \texttt{apdx376} (Appendix \ref{appendix-376-paper-558-evaluation-report}), particularly stood out
by leveraging the Chameleon Jupyter interface and providing a \emph{Trovi} \cite{noauthor_chameleoncloudtrovi_2025}
artifact — Chameleon’s platform for sharing reproducible artifacts. This
appendix’s artifact was reproducible with a single click. It serves as a prime
example of how straightforward artifact reproduction can be with well-prepared
tools and environment support, which sparked our interest to lower the barrier
of entry to the sophisticated automation systems provided by Chameleon Cloud.
This ultimately lead to our development of several \emph{Trovi} templates that
authors can utilize to more easily automate their artifacts as described in Section \ref{sec:trovi}.

\section{Results}

This section presents the outcomes of our evaluation and the key takeaways from the process. We distill these findings into guidelines and best practices for improving artifact reproducibility. In addition, we provide an overview of the new \emph{Trovi} templates developed during the project.

\subsection{Artifact Evaluation Results}

Table \ref{tab:ae} summarizes the results of our artifact evaluation. For each
appendix considered for a ``Results Replicated'' badge, we assign a score from 1
to 3 (3 being best) reflecting how well the artifact could be reproduced and how
many issues were encountered. A score of 3 means no issues were encountered; 2
indicates minor issues; and 1 indicates major issues that required significant
effort to resolve.

\begin{table}[h]
  \caption{Artifact Evaluation Results.}
  \label{tab:ae}
  \centering
  \begin{tabular}{lccc}
   \toprule
   \bf ID & \bf Reviewer verdict & \bf Own verdict & \bf Score \\
   \midrule
111 & RR & RR & 2\\
174 & AF & AF & -\\
190 & RR & RR & 2\\
\rowcolor{Gray} 191 & RR & AF & -\\
202 & RR & RR & 1\\
227 & AA & AA & -\\
332 & RR & RR & 3\\
\rowcolor{Gray} 362 & AF & RR & 3\\
368 & RR & RR & 3\\
376 & RR & RR & 3\\
\rowcolor{Gray} 407 & AF & RR & 1\\
428 & AA & AA & -\\
\rowcolor{Gray} 466 & AF & RR & 1\\
467 & RR & RR & 3\\
482 & AA & AA & -\\
\rowcolor{Gray} 483 & AF & RR & 2\\
496 & RR & RR & 2\\
506 & RR & AA & 1\\
    \bottomrule
    \end{tabular}
\end{table}

In Table \ref{tab:ae}, \textbf{RR} denotes a ``Results Replicated'' verdict,
\textbf{AF} denotes ``Artifacts Evaluated -- Functional'', and \textbf{AA} denotes
``Artifacts Available.'' The table shows both the official reviewer’s verdict
and our own verdict for each appendix, we highlight rows where our verdict
diverged from the reviewer's verdict.

For most appendices, our verdict aligned with the reviewer’s verdict. However,
determining what qualifies as a ``Results Replicated'' outcome can be somewhat
subjective. Different hardware or experimental setups naturally lead to slightly
different outcomes, and without clearly stated expected results in the
appendices, it is difficult to judge reproducibility -- particularly for
domain-specific results. Defining a precise boundary between an \textbf{AF} and an \textbf{RR}
badge is challenging, and even the official reviewers sometimes struggled with
this distinction.

Another area for improvement is clarifying the definition of an ``artifact'' in
this context. The template for artifact evaluation appendices suggests that each
\emph{artifact} corresponds to specific contributions of the
paper, suggesting that each should be a self-contained section of the appendix.
In practice, however, many authors misinterpret this and use multiple
artifact entries purely for setting up software or downloading data. We suggest
the Reproducibility Committee providing a clearer definition of what constitutes
an artifact and encouraging authors to structure their appendices accordingly.
Furthermore, we see potential for a dedicated \emph{software setup} section
so authors have a streamlined way to describe the necessary steps to create
an environment their experiments can be executed in.

Based on these findings, the following subsections offer guidelines and best
practices for authors to improve the reproducibility of their computational
artifacts.

\subsection{Guidelines for Authors}
\begin{enumerate}
\item \textbf{Documentation}: Provide clear and detailed documentation in the Artifact Evaluation appendix so that the artifact can be reproduced without relying on external resources. For third-party software, it is acceptable to refer to official documentation as needed.
\item \textbf{Software}: Clearly specify the versions of all required software components used (operating system, libraries, tools). Also, enumerate all steps needed to set up the software environment.
\item \textbf{Hardware}: State the hardware on which the experiment was conducted. In particular, note the architecture the experiments are intended for, and ensure that any provided software (e.g., Docker images) is compatible with commonly available architectures.
\item \textbf{Experiment Data}: Publish the experiment data to a public repository (e.g., Zenodo) and ensure it remains accessible to reviewers and readers during the evaluation period.
\item \textbf{Expected Results}: Describe the expected outcomes of the experiments in the appendix, especially if the reviewer’s hardware or environment differs from what was originally used.
\item \textbf{Long Experiments}: For long-running experiments, provide periodic progress output so the reviewer knows the experiment is proceeding as expected. In the documentation, indicate
\begin{enumerate}[label=(\alph*)]
\item the approximate duration of each long-running step, and
\item how progress is reported (for example, the frequency of status messages or a sample of the output).
\end{enumerate}
\item \textbf{Sample Execution}: Perform a trial run of your artifact using the provided setup instructions to ensure it executes as expected from start to finish.
\end{enumerate}

\subsection{Best Practice Recommendations}
\begin{enumerate}
\item \textbf{Reproducible Environment}: Use a reproducible environment for the artifact. This can be achieved in several ways:
  \begin{enumerate}[label=(\alph*)]
  \item \textbf{Containerization Solutions}: Provide instructions to build the environment or, ideally, supply a ready-to-use container image. For example, Docker, Singularity, or VirtualBox images can encapsulate the entire runtime environment.
  \item \textbf{Reproducible Builds}: Consider using package managers like Nix \cite{dolstra_purely_2006} or Guix \cite{courtes_functional_2013}, which have gained popularity for creating reproducible environments with exact software versions across different systems.
  \end{enumerate}
\item \textbf{Partial Automation}: Break complex experiments into smaller, manageable steps (e.g., separate shell scripts for setup, execution, analysis). Prefix each script with a number to clearly indicate the execution order.
\item \textbf{X11 Availability}: Reviewers typically do not have access to a graphical user interface on the evaluation system. If your artifact requires a GUI, provide a headless mode to run it. For instance, save plots or other visual output to files instead of opening an interactive window (e.g., avoid using \texttt{plt.show()} without an X11 display).
\item \textbf{Experiment Output}: Avoid including pre-generated output files in the artifact unless they are specifically meant for comparison or validation. If such output files are provided, clearly label them (e.g., include “expected” in the filename). Similarly, do not ship logs or notebook outputs as part of the artifact – these should be generated fresh during the artifact evaluation.
\end{enumerate}

\subsection{Trovi Templates}
\label{sec:trovi}

Trovi already offers some artifact templates, and we expanded this catalog by contributing three new templates. All our templates share a common structure (including a \texttt{config.py} for central configuration of parameters to provision a node and run the experiments) and have been published to Trovi’s public catalog.

\begin{enumerate}
\item \textbf{Docker Template}: If an NVIDIA GPU is available on the system, this template installs the NVIDIA Container Toolkit and then runs a sample operation inside a Docker container. The output of that operation is copied back to the host system for verification.
\item \textbf{Nix Template}: Installs the Nix package manager on the provisioned node and launches a Nix shell environment with the Julia programming language. The template then uses Julia to compute an 800×800 image of the Mandelbrot set as a demonstration.
\item \textbf{Guix Template}: Installs the Guix package manager on the node and runs the artifacts from Courtès et al.~\cite{courtes_storage_2006}, which were adapted to Guix in 2020~\cite{courtes_ten_2020}.
\end{enumerate}

\bibliographystyle{plain}
\bibliography{references}

\pagebreak

\appendix
\section{Appendix 111 Evaluation
Report}\label{appendix-111-paper-220-evaluation-report}

\subsection{Artifact A1}\label{artifact-a1}

\subsubsection{Hardware and Software
Environment}\label{hardware-and-software-environment}

\begin{itemize}
\tightlist
\item
  \textbf{Location}: TACC
\item
  \textbf{Node type}: compute-skylake
\item
  \textbf{OS Image}: CC-Ubuntu22.04 (officially supported)
\item
  \textbf{CPUs}: 2
\item
  \textbf{Threads}: 48
\item
  \textbf{RAM Size}: 192 GiB
\end{itemize}

\subsubsection{Reproduction Attempt}\label{reproduction-attempt}

\begin{itemize}
\tightlist
\item
  \textbf{Date}: 2024-07-21
\item
  \textbf{Time}: 23:33
\end{itemize}

\subsubsection{Steps performed}\label{steps-performed}

\begin{enumerate}
\def\labelenumi{\arabic{enumi}.}
\tightlist
\item
  Followed the installation and deployment steps in the artifact
  evaluation appendix:

  \begin{itemize}
  \tightlist
  \item
    Cloned the repository
  \item
    Installed
    \texttt{cmake\ ninja-build\ build-essential\ libboost-all-dev\ libbz2-dev}
    from the apt package manager
  \item
    Built the project using \texttt{cmake\ -\/-preset\ Linux-Release} in
    the top-level directory and \texttt{cmake\ -\/-build\ .} in the
    \texttt{builds/Linux-Release} directory
  \end{itemize}
\item
  Executed the \texttt{adversary-sw-16.ini} configuration to test the
  simulator. Test was successful.
\item
  Wrote a script for executing all the configurations in a single batch
  unattended. Everything worked as expected and the results were
  organized in the \texttt{output} folder.
\end{enumerate}

\subsubsection{Notes}\label{notes}

\begin{itemize}
\tightlist
\item
  There are some issues with the provided configuration files. Firstly,
  the artifact evaluation appendix does not state which configurations
  belong to exactly which plot (only the group of plot they belong to),
  and the config file names are not very descriptive and use different
  nomenclature than the plots.
\item
  Automating the reproduction would be quite nice
\item
  No plotting scripts provided
\end{itemize}

\subsubsection{Verdict}\label{verdict}

The following table is a summary of the results obtained and re-plotted:

\begin{longtable}[]{@{}llll@{}}
\toprule\noalign{}
Figure & Configuration File & Reproducible & Notes \\
\midrule\noalign{}
\endhead
\bottomrule\noalign{}
\endlastfoot
~9 (b) & \texttt{intra-C-group-sw-16.ini.csv} & Yes & - \\
9 (b) & \texttt{intra-C-group-chiplet-16.ini.csv} & Yes & x4 \\
9 (c) & \texttt{local-sw-16.ini.csv} & ~Yes & - \\
9 (c) & \texttt{local-chiplet-16.ini.csv} & Yes & x4 \\
~10 (a) & ~\texttt{global-sw-16.ini.csv} & Yes & - \\
10 (b) & \texttt{global-chiplet-16.ini.csv} & Yes & ~x4~ \\
~11 (a) & \texttt{local-sw-32.ini.csv} & Yes & - \\
~11 (a) & \texttt{local-chiplet-32.ini.csv} & Yes & x8 \\
11 (b) & ~\texttt{global-sw-32.ini.csv} & Yes & - \\
11 (b) & ~\texttt{global-chiplet-32.ini.csv} & Yes & x8 \\
12 (a) & \texttt{hotspot-sw-16.ini.csv} & Yes & - \\
12 (a) & \texttt{hotspot-chiplet-16.ini.csv} & Yes & x4 \\
12 (b) & \texttt{adversary-sw-16.ini.csv} & Yes & - \\
12 (b) & \texttt{adversary-chiplet-16.ini.csv} & Yes & x8 \\
13 (b) & \texttt{allreduce-sw-16.ini.csv} & Yes & - \\
13 (b) & \texttt{allreduce-chiplet-16.ini.csv} & Yes & x4 \\
\end{longtable}

I suggest a script for automating the computation and organization of
the results and/or better description of what exactly is to be
replicated with the provided configuration files, and am sure that both
of these measures would lead to a more streamlined replication process.

Importantly, it was not clear enough from the AD/AE appendix that the
results from the chiplet runs have to be multiplied by the given factor
to match the results in the paper. Also, the inclusion of plotting
scripts would have been helpful for a more streamlined replication, as
comparing number is not always the most efficient way to verify the
results.

Since the software setup was very well-documented and the simulation ran
successfully with some results reproduced, an \textbf{Artifact
Replicable} verdict is appropriate regardless of the previously
described minor issues.

\subsection{Artifact A2}\label{artifact-a2}

\subsubsection{Hardware and Software
Environment}\label{hardware-and-software-environment-1}

\emph{Same as for A1}

\subsubsection{Reproduction Attempt}\label{reproduction-attempt-1}

\begin{itemize}
\tightlist
\item
  \textbf{Date}: 2024-07-22
\item
  \textbf{Time}: 0:33
\end{itemize}

\subsubsection{Steps performed}\label{steps-performed-1}

\begin{enumerate}
\def\labelenumi{\arabic{enumi}.}
\tightlist
\item
  The python script was executed to run all the calculations for Table
  III, no additional package installs were required
\item
  Results were saved to a text file and compared with the table in the
  paper
\end{enumerate}

\subsubsection{Notes}\label{notes-1}

\begin{itemize}
\tightlist
\item
  Stating a more specific (or minimum) python version rather than
  ``Python3'' would be beneficial, even if the script only contains
  simple calculations.
\end{itemize}

\subsubsection{Verdict}\label{verdict-1}

A couple of results did not match the table in the paper, but these were
very minor differences due to typos in the script and incorrect
rounding. See the following list: - Script output:
\texttt{SW-less\ Cable\ Length:\ 72887.07}, Paper: \texttt{72K}.
Incorrect rounding of the script output in the cable. Even if one
considers this as an intermediate result for the cable number
calculation, the rounding does not make sense. - Script output:
\texttt{Taper\ FT\ Cabinet\ Number:608.0}, Paper: \texttt{960}. This is
most likely a typo in the script and should read
\texttt{FT\ Cabinet\ Number} instead of
\texttt{Taper\ FT\ Cabinet\ Number}. Additionally,
\texttt{Taper\ FT\ Cabinet\ Number} appears twice in the script output.

I think these minor issues still allow for an \textbf{Artifact
Replicated} verdict.

\subsection{Overall Verdict}\label{overall-verdict}

With some minor issues in both artifacts, there is definitely room for
improvement in the artifact documentation and the scripts provided.
However, the results were mostly reproducible, and the software setup
was well-documented. Therefore, I suggest an \textbf{Artifact
Replicated} verdict for this appendix.

\section{Appendix 174 Evaluation
Report}\label{appendix-174-paper-236-evaluation-report}

\subsection{Artifact A1}\label{artifact-a1-1}

\subsubsection{Hardware and Software
Environment}\label{hardware-and-software-environment-2}

\begin{itemize}
\tightlist
\item
  \textbf{Location}: TACC
\item
  \textbf{Node type}: compute-skylake
\item
  \textbf{OS Image}: CC-Ubuntu22.04 (officially supported)
\item
  \textbf{CPUs}: 2
\item
  \textbf{Threads}: 48
\item
  \textbf{RAM Size}: 192 GiB
\end{itemize}

\subsubsection{Reproduction Attempt}\label{reproduction-attempt-2}

\begin{itemize}
\tightlist
\item
  \textbf{Date}: 2024-07-21
\item
  \textbf{Time}: 13:19
\end{itemize}

\subsubsection{Steps performed}\label{steps-performed-2}

\begin{enumerate}
\def\labelenumi{\arabic{enumi}.}
\tightlist
\item
  As the image is not available publicly, we used the Chameleon-provided
  \texttt{CC-Ubuntu22.04}image on a generic \texttt{compute-skylake}
  node at the TACC site.
\item
  Download artifacts from the provided link
\end{enumerate}

\begin{itemize}
\tightlist
\item
  \texttt{wget\ https://sandbox.zenodo.org/records/45066/files/nanding0701/\\workflow\_roofline\_scripts-v1.zip?download=1\ -O\ apdx174.zip}
\end{itemize}

\begin{enumerate}
\def\labelenumi{\arabic{enumi}.}
\setcounter{enumi}{2}
\tightlist
\item
  Install required software
\end{enumerate}

\begin{itemize}
\tightlist
\item
  \texttt{apt\ install\ unzip\ python3-pip} (unzip for artifacts, pip
  for python packages)
\end{itemize}

\begin{enumerate}
\def\labelenumi{\arabic{enumi}.}
\setcounter{enumi}{3}
\tightlist
\item
  unzip and cd into the artifact directory
\end{enumerate}

\begin{itemize}
\tightlist
\item
  \texttt{unzip\ apdx174.zip\ \&\ cd\ nanding0701-workflow\_roofline\_scripts-985c4e4}
\end{itemize}

\begin{enumerate}
\def\labelenumi{\arabic{enumi}.}
\setcounter{enumi}{4}
\tightlist
\item
  Set up python packages
\end{enumerate}

\begin{itemize}
\tightlist
\item
  \texttt{python3\ -m\ venv\ .venv;\ source\ .venv/bin/activate} (create
  and activate virtual environment)
\item
  \texttt{pip\ install\ matplotlib\ numpy} (install required packages)
\end{itemize}

\begin{enumerate}
\def\labelenumi{\arabic{enumi}.}
\setcounter{enumi}{5}
\tightlist
\item
  Remove all pdf files from the artifact and execute python files
\end{enumerate}

\begin{itemize}
\tightlist
\item
  \texttt{rm\ *.pdf}
\item
  \texttt{for\ f\ in\ *.py;\ do\ python3\ \$f;\ done}
\end{itemize}

\subsubsection{Notes}\label{notes-2}

\begin{itemize}
\tightlist
\item
  Where can I find the image?

  \begin{itemize}
  \tightlist
  \item
    I've searched the following (all) Chameleon Cloud repos, as the
    image stores are not shared:

    \begin{itemize}
    \tightlist
    \item
      TACC
    \item
      UC
    \item
      NU
    \item
      NCAR
    \item
      IIT
    \item
      EVL
    \item
      EDGE (Error fetching images)
    \end{itemize}
  \item
    It seems like the image is not publicly visible, but only within the
    project. Will start from scratch.
  \end{itemize}
\item
  \textbf{Python code quality}:

  \begin{itemize}
  \tightlist
  \item
    in the \texttt{example.py} file is not great. Not PEP-conforming,
    many lines of seemingly unused commented code.
  \item
    Code quality in more relevant files (starting with \texttt{WRF\_}),
    is okay.
  \item
    No \texttt{requirements.txt} provided.
  \item
    There is a lot of copy-pasted code between the plotting scripts,
    which could easily be avoided with better code organization.
  \end{itemize}
\end{itemize}

\subsubsection{Verdict}\label{verdict-2}

\begin{itemize}
\tightlist
\item
  Plots generated successfully, but slight discrepancies in the plots
  compared to the paper.
\item
  Note: The artifact evaluation appendix states that text labels are not
  present in the plots, so we will not consider this a failure. As all
  plots were apparently edited after generation, we will consider a plot
  as ``OK'' if the main features are present but note any discrepancies
  except missing text labels.
\end{itemize}

\begin{center}\rule{0.5\linewidth}{0.5pt}\end{center}

\begin{longtable}[]{@{}
  >{\raggedright\arraybackslash}p{(\linewidth - 6\tabcolsep) * \real{0.2071}}
  >{\raggedright\arraybackslash}p{(\linewidth - 6\tabcolsep) * \real{0.4357}}
  >{\raggedright\arraybackslash}p{(\linewidth - 6\tabcolsep) * \real{0.1129}}
  >{\raggedright\arraybackslash}p{(\linewidth - 6\tabcolsep) * \real{0.2443}}@{}}
\toprule\noalign{}
\begin{minipage}[b]{\linewidth}\raggedright
Name
\end{minipage} & \begin{minipage}[b]{\linewidth}\raggedright
Corresponding Paper Figure
\end{minipage} & \begin{minipage}[b]{\linewidth}\raggedright
~Status
\end{minipage} & \begin{minipage}[b]{\linewidth}\raggedright
~ Comments
\end{minipage} \\
\midrule\noalign{}
\endhead
\bottomrule\noalign{}
\endlastfoot
~\texttt{example} & Figure 1 & OK & ~- \\
\texttt{WRF\_LCLS\_HSW} & Figure 5(a) & OK & ~(A), (B) \\
~\texttt{WRF\_LCLS\_PM} & Figure 6 & OK & ~(A), (B) \\
~\texttt{WRF\_BGW\_64} & Figure 7(a) & OK & ~(A), (C), (D) \\
~\texttt{WRF\_BGW\_1024} & Figure 7(b) & OK & ~(C), (D) \\
~\texttt{WRF\_BGW\_task} & Figure 7(c) & OK & ~(B), (E), both vertical
lines missing and inconsistent coloring of rooflines \\
~\texttt{WRF\_COSMO\_PM} & Figure 8 & OK & ~(B), rightmost vertical line
is on slightly different position (not as centered as in the paper
plot) \\
~\texttt{WRF\_GPTUNE\_PM} & Figure 10(a) & OK & ~(B), vertical line at 0
missing \\
\end{longtable}

Abbreviations: - (A): Colored areas not present - (B): Missing vertical
lines - (C): Dashed roofline not present - (D): Inconsistent z-level
(foreground/background) of plot elements - (E): Inconsistent colors

\textbf{All in all, the plots are okay, but the reproduction is by far
not perfect, and could be improved massively to more accurately reflect
the results from the paper. I am not sure if I would consider this
enough for a clean \texttt{Artifact\ Replicable} verdict}

\section{Appendix 190 Evaluation
Report}\label{appendix-190-paper-400-evaluation-report}

\subsection{Artifact A1}\label{artifact-a1-2}

\subsubsection{Hardware and Software
Environment}\label{hardware-and-software-environment-3}

\begin{itemize}
\tightlist
\item
  \textbf{Location}: TACC
\item
  \textbf{Node type}: compute-skylake
\item
  \textbf{OS Image}: CC-Ubuntu22.04 (officially supported)
\item
  \textbf{CPUs}: 2
\item
  \textbf{Threads}: 48
\item
  \textbf{RAM Size}: 192 GiB
\end{itemize}

\subsubsection{Reproduction Attempt}\label{reproduction-attempt-3}

\begin{itemize}
\tightlist
\item
  \textbf{Date}: 2024-07-22
\item
  \textbf{Time}: 18:31
\end{itemize}

\subsubsection{Steps performed}\label{steps-performed-3}

\begin{enumerate}
\def\labelenumi{\arabic{enumi}.}
\tightlist
\item
  Downloaded artifact from provided link (Zenodo) and unzipped it.
\item
  Evaluated the row count of the CSV files using
  \texttt{cat\ \$f\ \textbar{}~wc} and compared to the expected row
  count.
\end{enumerate}

\begin{itemize}
\tightlist
\item
  Command line input/output:
\end{itemize}

\begin{Shaded}
\begin{Highlighting}[]
\ExtensionTok{$}\NormalTok{ cd datasets}
\ExtensionTok{$}\NormalTok{ for x in }\PreprocessorTok{*}\NormalTok{.csv}\KeywordTok{;} \ControlFlowTok{do} \BuiltInTok{echo} \VariableTok{$x}\KeywordTok{;} \FunctionTok{cat} \VariableTok{$x} \KeywordTok{|} \FunctionTok{wc} \AttributeTok{{-}l}\KeywordTok{;} \ControlFlowTok{done}
\end{Highlighting}
\end{Shaded}

\begin{Shaded}
\begin{Highlighting}[]
\NormalTok{fugaku\_points.csv}
\NormalTok{9795168}
\NormalTok{powercontrols.csv}
\NormalTok{9484203}
\NormalTok{sysusage.csv}
\NormalTok{9484203}
\end{Highlighting}
\end{Shaded}

\subsubsection{Notes}\label{notes-3}

\begin{itemize}
\tightlist
\item
  The artifact production is not included here, only the length check. I
  am not sure whether the organization as a standalone artifact is
  necessary, this could have been included as a prerequisite in the main
  artifact.
\end{itemize}

\subsubsection{Verdict}\label{verdict-3}

The row count of the CSV files matches the expected row count, I guess
this is a successful reproduction.

\subsection{Artifact A2}\label{artifact-a2-1}

\subsubsection{Hardware and Software
Environment}\label{hardware-and-software-environment-4}

\emph{Same as for A1}

\subsubsection{Reproduction Attempt}\label{reproduction-attempt-4}

\begin{itemize}
\tightlist
\item
  \textbf{Date}: 2024-07-22
\item
  \textbf{Time}: 18:31
\end{itemize}

\subsubsection{Steps performed}\label{steps-performed-4}

\begin{enumerate}
\def\labelenumi{\arabic{enumi}.}
\tightlist
\item
  Installed python packages using a requirements.txt with the package
  version detail from the AE appendix.
\item
  Installed relevant R packages
\end{enumerate}

\begin{itemize}
\tightlist
\item
  On the clean Ubuntu install, this required a couple of additional
  system packages, which were prompted in a failed package installation
  (\texttt{tidyverse} was especially susceptible to this).
\end{itemize}

\begin{enumerate}
\def\labelenumi{\arabic{enumi}.}
\setcounter{enumi}{2}
\tightlist
\item
  Ran the R and python scripts in the \texttt{analysis} directory. See
  the relevant outputs (without messages/warnings) below
\end{enumerate}

\textbf{c1.R}: Ran without errors, no output.

\textbf{c1.py}:

\begin{verbatim}
Max power short: 272.0088
Max power long: 221.30598456433577
\end{verbatim}

\textbf{c2.R}:

\begin{verbatim}
[1] "TABLE I - POWER CONTROLS USAGE:"
  Boost Eco Retention jobspercentage groupspercentage averagepower
1   OFF OFF        ON         63.270            95.26         87.9
2    ON  ON        ON         12.416            26.78         73.2
3    ON  ON       OFF         10.285            14.45         55.7
4   OFF  ON       OFF          7.329             5.45         52.8
5   OFF  ON        ON          3.098            14.22         74.7
6   OFF OFF       OFF          2.963            27.73        101.3
7    ON OFF       OFF          0.269            12.32        112.3
8    ON OFF        ON          0.004             0.24         56.4
\end{verbatim}

\textbf{c3.R}
{
\small
\begin{verbatim}
       jid grp              pcmedt              pcmsdt        econm nnuma elp
1 21846566   1 2023-04-03 11:10:34 2023-04-03 11:09:06 80075.874431 27632  89
2 21846569   1 2023-04-03 13:00:52 2023-04-03 12:59:27 66757.128549 27632  86
3 21846573   1 2023-04-03 13:07:55 2023-04-03 13:07:32 27576.248799 26000  24
4 21846574   1 2023-04-03 13:49:14 2023-04-03 13:48:49 29371.630967 27632  25
5 21846575   1 2023-04-03 14:27:39 2023-04-03 14:27:14 21022.063602 27632  25
         econ2 usctmut   period  rscg jobtype wdt         BER
1 69921.835984 1769690 earning1 Other  Normal  42 OFF-OFF-OFF
2 67046.879028 1771010 earning1 Other  Normal  27 OFF-OFF-OFF
3 15828.320736 1619700 earning1 Other  Normal  38 OFF-OFF-OFF
4 16978.297112 1773510 earning1 Other  Normal  39 OFF-OFF-OFF
5 17769.333856 1775260 earning1 Other  Normal  26 OFF-OFF-OFF

[1] "TABLE IV - WAIT TIME IN PRIORITY QUEUE:"
[1] "1st redeeming:"
[1] "Priority:"
# A tibble: 3 × 2
  jobtype average_wait_hours
  <chr>                <dbl>
1 Bulk                  0.48
2 Normal                0.97
3 Step                  2.86
[1] "Non priority:"
# A tibble: 3 × 2
  jobtype average_wait_hours
  <chr>                <dbl>
1 Bulk                  35.9
2 Normal                23.3
3 Step                 164.
[1] "2nd redeeming:"
[1] "Priority:"
# A tibble: 3 × 2
  jobtype average_wait_hours
  <chr>                <dbl>
1 Bulk                  0.12
2 Normal                0.31
3 Step                 81.2
[1] "Non priority:"
# A tibble: 3 × 2
  jobtype average_wait_hours
  <chr>                <dbl>
1 Bulk                  8.39
2 Normal                9.39
3 Step                160.
\end{verbatim}
}

\subsubsection{Notes}\label{notes-4}

\begin{itemize}
\tightlist
\item
  During execution of all scripts, there were quite a few warnings and
  messages.

  \begin{itemize}
  \tightlist
  \item
    In the python script particularly, even with the correct package
    versions, there were deprecation warnings from pandas. This could
    easily be avoided with better code maintenance.
  \item
    In R, the warnings were mostly masking warnings and one deprecation
    warning from \texttt{ggplot2}.
  \end{itemize}
\end{itemize}

\subsubsection{Verdict}\label{verdict-4}

All figures were successfully reproduced and look identical to the ones
in the paper. The produced text files were checked using \texttt{diff}
and were identical to the reference files from the artifact repository.

\subsection{Overall Verdict}\label{overall-verdict-1}

This is a very clear \textbf{Artifact Reproducible}. The installation
process of R and python packages could have been a little more
streamlined and well explained with some reference image on Chameleon.
For example, a hint on which \texttt{apt} packages are required would
have been helpful, and is definitely something that could be included in
guidelines for future artifacts.

\section{Appendix 191 Evaluation
Report}\label{appendix-191-paper-169-evaluation-report}

\subsection{Artifact A1}\label{artifact-a1-3}

\subsubsection{Hardware and Software
Environment}\label{hardware-and-software-environment-5}

\begin{itemize}
\tightlist
\item
  \textbf{Location}: TACC
\item
  \textbf{Node type}: icelake\_r650
\item
  \textbf{OS Image}: CC-Ubuntu22.04 (officially supported)
\item
  \textbf{CPUs}: 2
\item
  \textbf{Threads}: 160
\item
  \textbf{RAM Size}: 256 GiB
\end{itemize}

\subsubsection{Reproduction Attempt}\label{reproduction-attempt-5}

\begin{itemize}
\tightlist
\item
  \textbf{Date}: 2024-08-05
\item
  \textbf{Time}: 13:23
\end{itemize}

\subsubsection{Steps performed}\label{steps-performed-5}

\begin{enumerate}
\def\labelenumi{\arabic{enumi}.}
\tightlist
\item
  Artifact Setup:

  \begin{itemize}
  \tightlist
  \item
    Installed Singularity following the online documentation.
  \item
    Downloaded the pre-built Singularity image from the provided link.
  \item
    Built the image file using the provided data
  \end{itemize}
\item
  Running \texttt{/home/nyx256/go.sh} in the Singularity container
  fails, installing MPI in the docker container.
\item
  As pointed out in the other review, OpenMPI is needed, install from
  \url{https://download.open-mpi.org/nightly/open-mpi/v4.1.x/openmpi-v4.1.x-202403270313-6a7ee37.tar.bz2},
  then continued using the following guide:
  \url{https://edu.itp.phys.ethz.ch/hs12/programming\_techniques/openmpi.pdf}.
  Also, setting \texttt{OMPI\_ALLOW\_RUN\_AS\_ROOT=1} and
  \texttt{OMPI\_ALLOW\_RUN\_AS\_ROOT\_CONFIRM=1} is required.
\item
  After installing OpenMPI, the \texttt{/home/nyx256/go.sh} script runs
  without any issues and terminated after a few (\textless5) minutes.
\item
  Ran \texttt{/home/nyx256/otfile/io.sh} with the following output
\end{enumerate}

\begin{verbatim}
*************** Error Bound 1 ***************
---------- Writing Time for Ours ----------
Ours Preprocess time = 0.28 seconds
Ours Compression+Writing time = 1.73 seconds
Ours Total time = 2.03 seconds

---------- Writing Time for AMRIC ---------
AMRIC Preprocess time = 0.90 seconds
AMRIC Compression+Writing time = 1.54 seconds
AMRIC Total time = 2.40 seconds

*************** Error Bound 2 ***************
---------- Writing Time for Ours ----------
Ours Preprocess time = 0.29 seconds
Ours Compression+Writing time = 1.72 seconds
Ours Total time = 2.02 seconds

---------- Writing Time for AMRIC ---------
AMRIC Preprocess time = 0.93 seconds
AMRIC Compression+Writing time = 1.58 seconds
AMRIC Total time = 2.44 seconds
\end{verbatim}

\begin{verbatim}
- Output matches expected results (+/- 20ms)
\end{verbatim}

\begin{enumerate}
\def\labelenumi{\arabic{enumi}.}
\setcounter{enumi}{5}
\tightlist
\item
  Ran \texttt{run/decomp.sh} and \texttt{run/qualityCR.sh}:
\end{enumerate}

\begin{verbatim}
************************* Error bound 1 *************************
---------- Data Quality and CR for Baseline-SZ3 ----------
PSNR: 54.56 | Compression Ratio: 165.22
---------- Data Quality and CR for AMRIC-SZ3 -------------
PSNR: 54.83 | Compression Ratio: 200.96
---------- Data Quality and CR for Ours(pad+eb) ----------
PSNR: 60.62 | Compression Ratio: 302.75

************************* Error bound 2 *************************
---------- Data Quality and CR for Baseline-SZ3 ----------
PSNR: 56.67 | Compression Ratio: 142.06
---------- Data Quality and CR for AMRIC-SZ3 -------------
PSNR: 56.84 | Compression Ratio: 161.78
---------- Data Quality and CR for Ours(pad+eb) ----------
PSNR: 62.13 | Compression Ratio: 227.95
\end{verbatim}

\begin{verbatim}
- Output matches expected results exactly
\end{verbatim}

\begin{enumerate}
\def\labelenumi{\arabic{enumi}.}
\setcounter{enumi}{6}
\tightlist
\item
  \texttt{/home/wpx/go.sh} runs successfully and terminates after 3-4
  minutes.
\item
  Ran \texttt{/home/wpx/diags/decomp.sh} and
  \texttt{/home/wpx/diags/qualityCR.sh} with the following output:
\end{enumerate}

\begin{verbatim}
************************* Error bound 1 *************************
---------- Data Quality and CR for Baseline-SZ3 ----------
PSNR: 69.24 | Compression Ratio: 193.45
---------- Data Quality and CR for Ours(pad+eb) ----------
PSNR: 81.27 | Compression Ratio: 226.75

************************* Error bound 2 *************************
---------- Data Quality and CR for Baseline-SZ3 ----------
PSNR: 64.08 | Compression Ratio: 245.99
---------- Data Quality and CR for Ours(pad+eb) ----------
PSNR: 78.43 | Compression Ratio: 288.43
\end{verbatim}

\begin{verbatim}
- Output matches expected results exactly
\end{verbatim}

\begin{enumerate}
\def\labelenumi{\arabic{enumi}.}
\setcounter{enumi}{8}
\tightlist
\item
  Ran \texttt{/home/rtamr/go.sh}, terminated after \textasciitilde1
  minute.
\item
  Ran \texttt{/home/qualityCR.sh} with the following output:
\end{enumerate}

\begin{verbatim}
************************ Error bound 1 *************************
---------- Data Quality and CR for Baseline-SZ3 ---------
PSNR: 30.48 | Compression Ratio: 121.95
---------- Data Quality and CR for AMRIC-SZ3 ------------
PSNR: 30.06 | Compression Ratio: 114.08
---------- Data Quality and CR for TAC-SZ3 --------------
PSNR: 30.80 | Compression Ratio: 148.15
---------- Data Quality and CR for Ours(pad+eb) ---------
PSNR: 35.56 | Compression Ratio: 176.98

************************* Error bound 2 *************************
---------- Data Quality and CR for Baseline-SZ3 ---------
PSNR: 27.85 | Compression Ratio: 148.22
---------- Data Quality and CR for AMRIC-SZ3 ------------
PSNR: 27.49 | Compression Ratio: 149.78
---------- Data Quality and CR for TAC-SZ3 --------------
PSNR: 28.19 | Compression Ratio: 182.59
---------- Data Quality and CR for Ours(pad+eb) ---------
PSNR: 33.54 | Compression Ratio: 237.6
\end{verbatim}

\begin{verbatim}
- Output matches expected results exactly.
\end{verbatim}

\subsubsection{Notes}\label{notes-5}

\begin{itemize}
\tightlist
\item
  The authors use a singularity shell for dependency management and
  failed to include the requirement of OpenMPI. This could have been
  achieved with a pre-built image or clearer instructions.
\end{itemize}

\subsubsection{Verdict}\label{verdict-5}

The results are clearly reproducible for this appendix. Dependency
management nevertheless seems to be an overarching problem for some
appendices, as is in this one. Given that the authors communicated with
the reviewer and there were no other issues, this is a clear verdict of
\emph{Results Replicated}.

\subsection{Artifact A2}\label{artifact-a2-2}

\subsubsection{Hardware and Software
Environment}\label{hardware-and-software-environment-6}

\begin{itemize}
\tightlist
\item
  Same as in \emph{A1}
\end{itemize}

\subsubsection{Reproduction Attempt}\label{reproduction-attempt-6}

\begin{itemize}
\tightlist
\item
  \textbf{Date}: 2024-08-06
\item
  \textbf{Time}: 23:20
\end{itemize}

\subsubsection{Steps performed}\label{steps-performed-6}

\begin{enumerate}
\def\labelenumi{\arabic{enumi}.}
\tightlist
\item
  Similar to A1, spawned a singularity shell with the provided image
\item
  Ran \texttt{/home/post-mr/go.sh}. Run fails partially because of
  missing library \texttt{libzfp.so.1}.
\item
  Manually installed \texttt{zfp} library from source and added it to
  the \texttt{LD\_LIBRARY\_PATH}
\item
  Ran \texttt{/home/post-mr/go.sh}. Runs successfully in about one
  minute now.
\item
  Ran \texttt{/home/post-mr/qualityCR.sh}. Output:
\end{enumerate}

\begin{verbatim}
********** Data Quality and CR for ZFP **********
--------------- Error bound 1 --------------
Compression Ratio: 240
PSNR-Ori:  40.11
PSNR-Post: 42.19
--------------- Error bound 2 --------------
Compression Ratio: 147
PSNR-Ori: 43.67
PSNR-Post: 45.61

********** Data Quality and CR for SZ2 **********
--------------- Error bound 1 --------------
Compression Ratio: 170
PSNR-Ori:  41.87
PSNR-Post: 43.27
--------------- Error bound 2 --------------
Compression Ratio: 121
PSNR-Ori: 44.28
PSNR-Post: 45.93
\end{verbatim}

Results match expected results exactly.

\begin{enumerate}
\def\labelenumi{\arabic{enumi}.}
\setcounter{enumi}{3}
\tightlist
\item
  Ran \texttt{/home/post-uni/go.sh} (\textasciitilde1 minute)
\item
  Ran \texttt{/home/post-uni/time.sh}. Output:
\end{enumerate}

\begin{verbatim}
*************** Error Bound 1 ***************
----------Time of post-processing and ZFP (OpenMP)------
1. Time taken by I/O is : 0.954 sec
2. Time taken by Comp + DeComp is : 0.987 sec
3. Time taken by Sampling + Modeling is : 0.027 sec
4. Time taken by Process is : 0.109 sec
5. Time taken by Ori is : 1.941 sec
6. Time taken by Extra is : 0.136 sec
Overhead : 0.070

----------Time of post-processing and SZ2 (Serial) -----
1. Time taken by I/O is : 3.414 sec
2. Time taken by Comp + DeComp is : 4.909 sec
3. Time taken by Sampling + Modeling is : 0.026 sec
4. Time taken by Process is : 0.095 sec
5. Time taken by Ori is : 8.323 sec
6. Time taken by Extra is : 0.121 sec
Overhead : 0.015

*************** Error Bound 2 ***************
----------Time of post-processing and ZFP (OpenMP)------
1. Time taken by I/O is : 3.589 sec
2. Time taken by Comp + DeComp is : 0.933 sec
3. Time taken by Sampling + Modeling is : 0.022 sec
4. Time taken by Process is : 0.123 sec
5. Time taken by Ori is : 4.522 sec
6. Time taken by Extra is : 0.145 sec
Overhead : 0.032

----------Time of post-processing and SZ2 (Serial) -----
1. Time taken by I/O is : 3.438 sec
2. Time taken by Comp + DeComp is : 4.650 sec
3. Time taken by Sampling + Modeling is : 0.024 sec
4. Time taken by Process is : 0.086 sec
5. Time taken by Ori is : 8.088 sec
6. Time taken by Extra is : 0.109 sec
Overhead : 0.014
\end{verbatim}

Results match expected results (+/- 60ms) except Error Bound 1 for Ori (8.323 in reproduction vs. 5.790 expected)

\subsubsection{Notes}\label{notes-6}

\begin{itemize}
\tightlist
\item
  The not mentioned requirement of manually building a library for
  running the artifact is not very nice, and could definitely be
  improved by the authors. This is something that should at least be
  contained in a dedicated script instead of requiring manual building
  and installation of the library
\end{itemize}

\subsubsection{Verdict}\label{verdict-6}

This artifact is a little less documented than A1, but the results are
still mostly reproducible, which a second run of \texttt{go.sh}
\texttt{time.sh} in the last steps confirms. I think this still warrants
a \emph{Results Replicated} verdict.

\subsection{Overall Verdict}\label{overall-verdict-2}

While the artifacts are mostly reproducible, the lack of documentation
and the need for manual intervention in some cases is a bit of a
letdown. The authors should consider improving the documentation and the
scripts to make the artifacts more user-friendly. Nevertheless, the
results are mostly reproducible, and the artifacts are well-organized. I
would recommend a \emph{Results Replicated} verdict for this artifact.

\section{Appendix 202 Evaluation
Report}\label{appendix-202-paper-699-evaluation-report}

\subsection{Artifact A1}\label{artifact-a1-4}

\subsubsection{Hardware and Software
Environment}\label{hardware-and-software-environment-7}

\begin{itemize}
\tightlist
\item
  \textbf{Location}: TACC
\item
  \textbf{Node type}: icelake\_r650
\item
  \textbf{OS Image}: CC-Ubuntu22.04 (officially supported)
\item
  \textbf{CPUs}: 2
\item
  \textbf{Threads}: 160
\item
  \textbf{RAM Size}: 256 GiB
\end{itemize}

\subsubsection{Reproduction Attempt}\label{reproduction-attempt-7}

\begin{itemize}
\tightlist
\item
  \textbf{Date}: 2024-08-02
\item
  \textbf{Time}: 16:12
\end{itemize}

\subsubsection{Steps performed}\label{steps-performed-7}

\begin{enumerate}
\def\labelenumi{\arabic{enumi}.}
\tightlist
\item
  Installed \texttt{docker} using the instructions from
  \url{https://docs.docker.com/engine/install/ubuntu/}
\item
  Failed to pull image from \texttt{zhangwping/autocheck:latest}, docker
  hub reveals the image is called \texttt{zhangwping/vpautocheck}, which
  was then pulled.
\item
  Created container using
  \texttt{docker\ run\ -it\ zhangwping/vpautocheck:latest\ /bin/bash}
\item
  Executed \texttt{./run.sh} in the container in both configurations
  (Step 3 and 4 from the artifact evaluation appendix), which failed due
  to the missing \texttt{dot} executable. This only seems to be a small
  issue generating the figure, but the relevant files are still stored.
\item
  For Step 5 of the AE, the pre-built executable was used since the
  \texttt{cmake\ ..\ \&\&\ make} command failed due to the already
  existing build in the docker image.
\end{enumerate}

\begin{itemize}
\tightlist
\item
  Output of \texttt{checkpoint/Himeno/build/himeno\ 0}:
\end{itemize}

\begin{verbatim}
mimax = 129 mjmax = 65 mkmax = 65
imax = 128 jmax = 64 kmax =64
write check_0_0
Write tmp_0: the iter value is: 0
write check_0_1
Write tmp_0: the iter value is: 1
write check_0_2
Write tmp_0: the iter value is: 2
write check_0_3
Write tmp_0: the iter value is: 3
\end{verbatim}

\begin{itemize}
\tightlist
\item
  Output of \texttt{checkpoint/Himeno/build/himeno\ 1}:
\end{itemize}

\begin{verbatim}
imax = 129 mjmax = 65 mkmax = 65
imax = 128 jmax = 64 kmax =64
local_iter: 3
Read tmp_0: the iter value is: 3
read check_0_3
Rank 0 READCP file
write check_0_4
Write tmp_0: the iter value is: 4
write check_0_5
Write tmp_0: the iter value is: 5
write check_0_6
Write tmp_0: the iter value is: 6
write check_0_7
Write tmp_0: the iter value is: 7
write check_0_8
Write tmp_0: the iter value is: 8
write check_0_9
Write tmp_0: the iter value is: 9
cpu : 0.069578 sec.
Loop executed for 10 times
Gosa : 3.066880e-03
\end{verbatim}

\begin{itemize}
\tightlist
\item
  Output for \texttt{varify/Himeno/himeno}:
\end{itemize}

\begin{verbatim}
mimax = 129 mjmax = 65 mkmax = 65
imax = 128 jmax = 64 kmax =64
cpu : 0.036677 sec.
Loop executed for 10 times
Gosa : 3.066880e-03
MFLOPS measured : 4489.924476
Score based on MMX Pentium 200MHz : 139.136178
\end{verbatim}

\begin{itemize}
\tightlist
\item
  Except for the CPU number, the results between the second and third
  run match.
\end{itemize}

\subsubsection{Notes}\label{notes-7}

\begin{itemize}
\tightlist
\item
  It is very nice that the authors both provide an image and list the
  required libraries with versions.
\item
  Software configuration using interactive prompts is not ideal for
  reproducibility. Configuration files would be a better option.
\item
  There are some naming issues. For example, the executable is called
  \texttt{himeno}, not \texttt{Himeno}.
\end{itemize}

\subsubsection{Verdict}\label{verdict-7}

There is room for improvement in documentation and software setup. Also,
there could have been a better way to store artifacts to reproduce all
benchmarks, not only 2 selected ones, as these are not enough for
reproducing most results from the paper. For this reason, I am leaning
towards an \emph{Artifact functional} verdict.

\section{Appendix 227 Evaluation
Report}\label{appendix-227-paper-462-evaluation-report}

\subsection{Artifact A1}\label{artifact-a1-5}

\subsubsection{Hardware and Software
Environment}\label{hardware-and-software-environment-8}

\begin{itemize}
\tightlist
\item
  \textbf{Location}: UC
\item
  \textbf{Node type}: V100
\item
  \textbf{OS Image}: CC-Ubuntu22.04-CUDA (officially supported)
\item
  \textbf{CPUs}: 2
\item
  \textbf{Threads}: 80
\item
  \textbf{RAM Size}: 128 GiB
\item
  \textbf{GPU}: NVIDIA V100
\end{itemize}

\subsubsection{Reproduction Attempt}\label{reproduction-attempt-8}

\begin{itemize}
\tightlist
\item
  \textbf{Date}: 2024-08-18
\item
  \textbf{Time}: 22:36
\end{itemize}

\subsubsection{Steps performed}\label{steps-performed-8}

\begin{enumerate}
\def\labelenumi{\arabic{enumi}.}
\setcounter{enumi}{-1}
\tightlist
\item
  Downloaded the artifact from Zenodo, the rest of the evaluation is
  performed according to the \texttt{README.md} from the repository.
\item
  Built the \texttt{dm\_gen} executable in the \texttt{dm\_gen}
  directory using \texttt{make\ all\ \&\&\ make\ dm\_gen}. This required
  installing the \texttt{boost} library.
\item
  Cloned the \texttt{open-earth-compiler} and \texttt{llvm-project}
  repositories and applied the provided patch. This together with the
  following build step for \texttt{llvm} was consolidated into a script
  \texttt{02\_clone\_build\_llvm.sh}.
\item
  Built \texttt{open-earth-compiler}, following the provided steps and
  wrote it to the script \texttt{03\_build\_openearth.sh}. After adding
  \texttt{/usr/local/cuda/include} to the \texttt{CPLUS\_INCLUDE\_PATH}
  environment variable, the build and subsequent tests were successful.
\item
  Building the relay and fused stencil generation were skipped, as they
  are provided in the artifact and described as an optional steps
\item
  Tried running the \texttt{run.py} script in the \texttt{ad/artemis}
  directory. This failed due to missing dependencies, which were not
  specified in the README, namely \texttt{tvm}. After installing
  \texttt{pip} and \texttt{apache-tvm}, there were issues due to an
  incompatible numpy version, as \texttt{np.float\_} is deprecated since
  version 2.0. It turns out that building Apache TVM is indeed necessary
  due to modifications by the authors.
\item
  Built the provided Apache TVM version \texttt{tvm-graph}. This
  required the \texttt{sympy}, \texttt{tqdm}, \texttt{torch}, and
  \texttt{scikit-learn} packages, which were again not listed as
  dependencies.
\item
  After building, the experiment did again not run due to including a
  \texttt{common.hpp} file. This could only be resolved with a symbolic
  link from the \texttt{/workspace} directory (thanks to the
  reproducibility reviewers for this suggestion!) to the actual path of
  the \texttt{common.hpp} file.
\item
  Now, \texttt{ad/moirae/run.py} could be executed, but does not provide
  any output, neither do the other \texttt{run.py} scripts. Neither of
  the scripts terminates after 60 hours (this was repeated twice (once
  with 53, once with 60 hours left on the lease) on two different nodes,
  but the reservation of the V100 node could not be extended further in
  both cases).
\end{enumerate}

\subsubsection{Notes}\label{notes-8}

\begin{itemize}
\tightlist
\item
  There is no Artifact Evaluation Appendix present. Instead, the
  artifact description references the \texttt{README} files from the
  artifact repository. Technically, this would disqualify the artifact
  from being evaluated.
\item
  The authors did not state any dependency versions, neither in the
  artifact nor in the README.
\item
  The running time of the artifact is incredibly frustrating, as there
  is no progress output whatsoever.
\end{itemize}

\subsubsection{Verdict}\label{verdict-8}

Between the arduous build process which took more than 4 hours, the
missing dependencies, and the advertised running time that exceeds the 8
hours by far, the artifact is definitely not reproducible. I would even
lean towards a verdict of \texttt{Artifact\ Available} since there is no
Artifact Evaluation Appendix present and there are numerous,
undocumented, errors in the build process.

\section{Appendix 332 Evaluation
Report}\label{appendix-332-paper-594-evaluation-report}

\subsection{Artifact A1}\label{artifact-a1-6}

\subsubsection{Hardware and Software
Environment}\label{hardware-and-software-environment-9}

\begin{itemize}
\tightlist
\item
  \textbf{Location}: UC
\item
  \textbf{Node type}: gtx6000
\item
  \textbf{OS Image}: CC-Ubuntu22.04 (officially supported)
\item
  \textbf{CPUs}: 2
\item
  \textbf{Threads}: 48
\item
  \textbf{RAM Size}: 192 GiB
\end{itemize}

\subsubsection{Reproduction Attempt}\label{reproduction-attempt-9}

\begin{itemize}
\tightlist
\item
  \textbf{Date}: 2024-08-06
\item
  \textbf{Time}: 02:44
\end{itemize}

\subsubsection{Steps performed}\label{steps-performed-9}

Followed the very detailed instructions in the appendix: 1. Downloaded
artifact from \url{https://github.com/IntelliSys-Lab/Stellaris-SC24}
2. Installed \texttt{docker}and NVIDIA drivers using the provided
scripts in \texttt{evaluation} folder 3. Pulled docker image 4. Ran
experiments using the \texttt{run\_experiments.sh} script - Seems to be
stuck on \texttt{Connected\ to\ Ray\ cluster.} Did not change in the
given time limit of 20 minutes. - Output:
\texttt{2024-08-06\ 01:33:59,780\ INFO\ worker.py:1490\ -\/-\ Connecting\ to\ existing\ Ray\ cluster\ at\ address:\ 172.18.0.2:6380...}
- Script was terminated after 30 minutes with no changes, retry in a
tmux session and let it run overnight - Update after 1 hour of running
time: Output still stuck at same point, but htop shows there is progress
happening 5. After not terminating the first time, the experiment was
replicated again, this time with progress output and successful
termination. 6. The figures compare well with the ones in the paper,
with minor differences in visual appearance, but capturing the same
trends.

\subsubsection{Notes}\label{notes-9}

\begin{itemize}
\tightlist
\item
  While the description is very nice and detailed, the dependencies
  could have been included in a docker image, especially since this
  software setup is common and relevant to other Chameleon users.
\item
  It is incredibly nice for arduous installation steps to be automated
  using scripts
\item
  \textbf{IDEA FOR GUIDELINES: take some examples from here (scripts,
  automated dependency install or provided image)}
\item
  \textbf{IDEA FOR BEST-PRACTICE GUIDELINES: Show progress in
  long-running applications. Otherwise, the process is incredibly
  frustrating}
\end{itemize}

\subsubsection{Verdict}\label{verdict-9}

With a very clear, easily reproducible appendix, this is the most clear
\emph{results replicated} yet without question.

\section{Appendix 362 Evaluation
Report}\label{appendix-362-paper-443-evaluation-report}

\subsection{Artifact A1}\label{artifact-a1-7}

\subsubsection{Hardware and Software
Environment}\label{hardware-and-software-environment-10}

\begin{itemize}
\tightlist
\item
  \textbf{Location}: UC
\item
  \textbf{Node type}: gtx6000
\item
  \textbf{OS Image}: CC-Ubuntu22.04-CUDA (officially supported)
\item
  \textbf{CPUs}: 2
\item
  \textbf{Threads}: 48
\item
  \textbf{RAM Size}: 192 GiB
\item
  \textbf{GPU}: 1x Nvidia GTX 6000
\end{itemize}

\subsubsection{Reproduction Attempt}\label{reproduction-attempt-10}

\begin{itemize}
\tightlist
\item
  \textbf{Date}: 2024-08-07
\item
  \textbf{Time}: 19:50
\end{itemize}

\subsubsection{Steps performed}\label{steps-performed-10}

For the provisioned machine, a Chameleon Share (NFS) was created and
attached for the large amount of experiment data.

\begin{itemize}
\tightlist
\item
  Installed docker using the official instructions on the docker
  website. For this, a script \texttt{install\_docker.sh} was created
  that also adds the current user to the docker group.
\item
  Installed the nvidia container toolkit using the script from the
  appendix. This was also extracted to a script
  \texttt{install\_nvidia\_container\_toolkit.sh}.
\item
  Pulled the docker image from Dockerhub and ran a bash shell inside the
  container to confirm the image was working.
\end{itemize}

\subsubsection{Notes}\label{notes-10}

\begin{itemize}
\tightlist
\item
  This issue has already occurred in other implementations, but an
  artifact consisting solely of a docker container does not really
  contribute to any results from the paper. This is something that
  should be included in the guidelines for future artifacts.
\end{itemize}

\subsubsection{Verdict}\label{verdict-10}

Since this artifact only includes the successful download of the docker
container, the reproduction was successful. However, the artifact does
not represent any contribution to the paper.

\subsection{Artifacts A2 and A3}\label{artifacts-a2-and-a3}

\subsubsection{Hardware and Software
Environment}\label{hardware-and-software-environment-11}

\textbf{Same as A1}

\subsubsection{Reproduction Attempt}\label{reproduction-attempt-11}

\textbf{Same as A1}

\subsubsection{Steps performed}\label{steps-performed-11}

\begin{enumerate}
\def\labelenumi{\arabic{enumi}.}
\tightlist
\item
  Downloaded the artifact from Zenodo using the provided script and
  extracted the contents.
\item
  Compared the GPU- and CPU-generated data visually. Besides mirroring
  and reordering of certain parts of the chormosomes, the images are
  very similar for all generated chromosome layouts.
\item
  Furthermore, the figures were compared to the ones from ``Pangenome
  graph layout by Path-Guided Stochastic Gradient Descent'' paper by
  Heumos et al.
\end{enumerate}

\subsubsection{Notes}\label{notes-11}

\begin{itemize}
\tightlist
\item
  Again, the contributions linked to these artifacts are not significant
  enough to warrant the organization as a separate artifact.
\end{itemize}

\subsubsection{Verdict}\label{verdict-11}

The download of the data was successful, but it remains unclear what the
contribution of these artifacts is to the paper except providing a
baseline. This should really be included in a potential software setup
section. Nevertheless, this can be considered a successful reproduction
according to the author's instructions.

\subsection{Artifacts A4 and A5}\label{artifacts-a4-and-a5}

\subsubsection{Hardware and Software
Environment}\label{hardware-and-software-environment-12}

\textbf{Same as A1}

\subsubsection{Reproduction Attempt}\label{reproduction-attempt-12}

\textbf{Same as A1}

\subsubsection{Steps performed}\label{steps-performed-12}

\begin{enumerate}
\def\labelenumi{\arabic{enumi}.}
\tightlist
\item
  Cloned the \texttt{tonyjie/gpu\_pangenome\_layout\_artifact}
  repository
\item
  Ran the docker image, attaching the zenodo artifact and the repository
  as volumnes.
\item
  Downloaded the dataset using \texttt{dataset\_preprocess.sh}.
  Especially the layout generation using \texttt{ogdi} exceeded the
  stated time period of 1-2 hours, taking about 3 hours to complete on
  the provided machine.
\item
  Ran the GPU experiments using \texttt{run\_gpu\_layout\_all.sh}. This
  step took 193 minutes to complete.
\item
  Ran the CPU experiments using \texttt{run\_cpu\_layout\_all.sh}. This
  step was run over two and a half days, resulting in a completion time
  of a less than 36 hours (exact time reporting was not written to the
  intended log file).
\item
  Compared the results of the GPU and CPU experiments using\\
  \texttt{run\_path\_stress\_verify.sh}.
\end{enumerate}

\subsubsection{Notes}\label{notes-12}

\begin{itemize}
\tightlist
\item
  The long-running scripts output progress updates, which is a good
  practice for long-running experiments. Nevertheless, some commands
  could be more verbose to indicate progress.
\end{itemize}

\subsubsection{Verdict}\label{verdict-12}

A clear speedup of the GPU over the CPU implementation was observed. The
reproduction was successful, and the results were as expected.

\subsubsection{Overall Verdict}\label{overall-verdict-3}

The reproduction was successful, and the results were as expected. While
this experiment technically exceeds the running time of the artifact
evaluation, the software setup was well-documented and no errors
occurred during the reproduction. A verdict of
\texttt{Results\ Replicated} is appropriate.

\section{Appendix 368 Evaluation
Report}\label{appendix-368-paper-386-evaluation-report}

\subsection{Artifact A1}\label{artifact-a1-8}

\subsubsection{Hardware and Software
Environment}\label{hardware-and-software-environment-13}

\begin{itemize}
\tightlist
\item
  \textbf{Location}: TACC
\item
  \textbf{Node type}: icelake\_r650
\item
  \textbf{OS Image}: CC-Ubuntu22.04 (officially supported)
\item
  \textbf{CPUs}: 2
\item
  \textbf{Threads}: 160
\item
  \textbf{RAM Size}: 256 GiB
\end{itemize}

\subsubsection{Reproduction Attempt}\label{reproduction-attempt-13}

\begin{itemize}
\tightlist
\item
  \textbf{Date}: 2024-08-16
\item
  \textbf{Time}: 22:46
\end{itemize}

\subsubsection{Steps performed}\label{steps-performed-13}

\begin{enumerate}
\def\labelenumi{\arabic{enumi}.}
\tightlist
\item
  Provisioned a new \texttt{icelake\_r650} node at TACC. Unfortunately,
  the image \texttt{kamping-ubuntu22.04-docker} is not available, likely
  due to a visibility misconfiguration.
\item
  Installed Docker according to the installation guide from the Docker
  website and added \texttt{cc} user to the \texttt{docker} group.
\item
  Executed script \texttt{00\_pull\_docker\_image.sh}, which pulls the
  Docker image \texttt{kamping-site/kamping-reproducibility} from Docker
  Hub.
\item
  Executed script \texttt{01\_run\_docker\_container\_here.sh}, entering
  the docker container.
\item
  Cloned the repository \texttt{kamping-site/kamping} from GitHub and
  checked out branch \texttt{v0.0.1}.
\item
  Built the project using \texttt{cmake} as instructed.
\item
  Executed all tests using \texttt{ctest}, all tests passed.
  \texttt{100\%\ tests\ passed,\ 0\ tests\ failed\ out\ of\ 551}
\end{enumerate}

\subsubsection{Notes}\label{notes-13}

\begin{itemize}
\tightlist
\item
  It is a nice idea of the authors to leverage the template such that
  the first section is dedicated to general software setup. It may be
  beneficial to include this in the guidelines or suggest a template
  adaptation.
\item
  There is a \texttt{\textbackslash{}} missing in the
  \texttt{git\ clone} command from step 5.
\end{itemize}

\subsubsection{Verdict}\label{verdict-13}

The software built successfully and all tests passed, making this
artifact reproducible. Normally, software setup does not count as an
artifact, but in this case, the successful compilation of the software
is directly relevant to the paper content (Figures 4-6)

\subsection{Artifact A2}\label{artifact-a2-3}

\subsubsection{Hardware and Software
Environment}\label{hardware-and-software-environment-14}

\textbf{Same as A1}

\subsubsection{Reproduction Attempt}\label{reproduction-attempt-14}

\textbf{Same as A1}

\subsubsection{Steps performed}\label{steps-performed-14}

\begin{enumerate}
\def\labelenumi{\arabic{enumi}.}
\tightlist
\item
  Cloned the \texttt{kamping-site/kamping-examples} repository from
  GitHub and checked out branch \texttt{v0.1.1} using the provided
  script \texttt{00\_fetch\_source.sh}.
\item
  Built the experiments using \texttt{01\_build\_benchmarks.sh}
\item
  Ran \texttt{02\_count\_loc.sh}, producing the following output, that
  matches the results from the paper:
\end{enumerate}

\begin{verbatim}
application,implementation,length,begin,end,comment_lines,blank_lines
BFS,BOOST,42,7,52,1,1
BFS,KAMPING,31,7,40,0,1
BFS,KAMPING_FLATTENED,22,8,32,0,1
BFS,KAMPING_GRID,24,7,33,0,1
BFS,KAMPING_SPARSE,28,9,39,0,1
BFS,MPI,46,8,56,0,1
BFS,MPI_NEIGHBORHOOD,54,8,64,0,1
BFS,MPI_NEIGHBORHOOD_DYNAMIC,55,8,66,0,2
BFS,MPL,49,6,57,0,1
BFS,RWTH_MPI,32,7,41,0,1
SORTING,BOOST,30,7,39,1,0
SORTING,KAMPING,28,10,40,1,0
SORTING,KAMPING_FLATTENED,16,10,27,0,0
SORTING,MPI,32,8,42,1,0
SORTING,MPL,37,7,45,0,0
SORTING,RWTH_MPI,21,9,31,0,0
VECTOR_ALLGATHER,BOOST,5,9,15,0,0
VECTOR_ALLGATHER,KAMPING,1,6,8,0,0
VECTOR_ALLGATHER,MPI,14,10,25,0,0
VECTOR_ALLGATHER,MPL,12,6,19,0,0
VECTOR_ALLGATHER,RWTH_MPI,5,12,18,0,0
\end{verbatim}

\begin{enumerate}
\def\labelenumi{\arabic{enumi}.}
\setcounter{enumi}{3}
\tightlist
\item
  Ran \texttt{03\_run\_sorting.sh}, 0 out of 48 runs failed.
\item
  Ran \texttt{04\_run\_bfs.sh}, 0 out of 256 runs failed.
\item
  Ran \texttt{05\_eval\_sorting.sh}, writing the output to
  \texttt{sorting.csv} and creating the figure \texttt{sorting.pdf}.
\item
  Ran \texttt{06\_eval\_bfs.sh}, writing the output to \texttt{bfs.csv}
  and creating the figure \texttt{bfs.pdf}.
\end{enumerate}

The created plots roughly match the ones from the paper. Considering the
test architecture is different, the results are close enough. Namely,
the scalability of kaMPIng is not as visible in the reproduced plots as
in the paper.

\subsubsection{Notes}\label{notes-14}

\begin{itemize}
\tightlist
\item
  In step 3, the CSV output column names do not match the ones from the
  paper exactly, this could be improved for clarity.
\end{itemize}

\subsubsection{Verdict}\label{verdict-14}

Considering the results are close enough to the ones from the paper,
this artifact is reproducible. The authors should consider improving the
CSV output column names for clarity.

\subsection{Artifact A3}\label{artifact-a3}

\subsubsection{Hardware and Software
Environment}\label{hardware-and-software-environment-15}

\textbf{Same as A1}

\subsubsection{Reproduction Attempt}\label{reproduction-attempt-15}

\textbf{Same as A1}

\subsubsection{Steps performed}\label{steps-performed-15}

\begin{enumerate}
\def\labelenumi{\arabic{enumi}.}
\tightlist
\item
  Cloned the respective repository using the provided script
  \texttt{00\_fetch\_source.sh}.
\item
  Counted the lines of code using \texttt{01\_count\_loc.sh}.
\end{enumerate}

The lines of code reported in the paper match the ones from the
reproduced output exactly.

\subsubsection{Notes}\label{notes-15}

The output could have been structured better to match the results from
the paper more closely.

\subsubsection{Verdict}\label{verdict-15}

The artifact is clearly reproducible.

\subsection{Artifact A4}\label{artifact-a4}

\subsubsection{Hardware and Software
Environment}\label{hardware-and-software-environment-16}

\textbf{Same as A1}

\subsubsection{Reproduction Attempt}\label{reproduction-attempt-16}

\textbf{Same as A1}

\subsubsection{Steps performed}\label{steps-performed-16}

\begin{enumerate}
\def\labelenumi{\arabic{enumi}.}
\tightlist
\item
  Cloned the respective repository using the provided script
  \texttt{00\_fetch\_source.sh}.
\item
  Built the benchmark using \texttt{01\_build\_benchmark.sh}.
\item
  Counted the lines of code using \texttt{02\_count\_loc.sh}. Again, the
  results match the ones from the paper exactly.
\item
  Ran the benchmarks using \texttt{03\_run\_benchmark.sh}. Finished
  without erros (0 out of 54 failed).
\item
  Evaluated the results using \texttt{04\_eval.sh}, producing the
  following output:
\end{enumerate}

\begin{verbatim}
PlainMPI:        1.966056
KaMPIngWrapper:      1.967056
dKaMinParWrapper:    2.015556
\end{verbatim}

\subsubsection{Notes}\label{notes-16}

\begin{itemize}
\tightlist
\item
  The AD/AE appendix does not state the expected output of the
  evaluation script.
\end{itemize}

\subsubsection{Verdict}\label{verdict-16}

Apart from the lines of code written in the paper, it is unclear which
results from the paper were reproduced by running the benchmarks. Since
however only the \emph{Graph Partitioning} section is stated to be
reproduced and everything ran as expected, the artifact can be
considered reproducible.

\subsection{Artifact A5}\label{artifact-a5}

\subsubsection{Hardware and Software
Environment}\label{hardware-and-software-environment-17}

\textbf{Same as A1}

\subsubsection{Reproduction Attempt}\label{reproduction-attempt-17}

\textbf{Same as A1}

\subsubsection{Steps performed}\label{steps-performed-17}

\begin{enumerate}
\def\labelenumi{\arabic{enumi}.}
\tightlist
\item
  Cloned the respective repository using the provided script
  \texttt{00\_fetch\_source.sh}.
\item
  Built the benchmark using \texttt{01\_build\_benchmark.sh}. The
  compilation time increase with kaMPIng is around 18s (1:49 min vs 2:07
  min) or 20\%, matching the results from the paper.
\item
  Checked the binary size using \texttt{02\_binary\_size.sh}, producing
  the following output:
\end{enumerate}

\begin{verbatim}
Binary sizes:
RAxML-NG:
2.1M    raxml-ng/bin/raxml-ng-mpi
RAxML-NG + KaMPIng:
2.2M    kamping-raxml-ng/bin/raxml-ng-mpi
\end{verbatim}

\begin{enumerate}
\def\labelenumi{\arabic{enumi}.}
\setcounter{enumi}{3}
\tightlist
\item
  For more accurate results, the parameter of \texttt{du\ -h} was
  changed to \texttt{-b} in the script \texttt{02\_binary\_size.sh} to
  get the exact size in bytes. The output was:
\end{enumerate}

\begin{verbatim}
Binary sizes:
RAxML-NG:
2198736 raxml-ng/bin/raxml-ng-mpi
RAxML-NG + KaMPIng:
2254032 kamping-raxml-ng/bin/raxml-ng-mpi
\end{verbatim}

This output exactly matches the expected size increase of 2.5\% from the
paper.

\subsubsection{Notes}\label{notes-17}

\begin{itemize}
\tightlist
\item
  It is stated in the appendix that this benchmark is not intended to be
  run, as the experiment takes 5 hours and requires a large cluster.
\end{itemize}

\subsubsection{Verdict}\label{verdict-17}

As the authors explicitly excluded running the benchmark from the
reproducible artifacts, this artifact is reproducible as everything ran
as expected and the binary size matches the expected size from the
paper.

\subsection{Overall Verdict}\label{overall-verdict-4}

This is a well-prepared artifact, with a clear structure and detailed
instructions. Despite minor issues and slight discrepancies in benchmark
results, which are expected when changing system architecture, this
artifact is clearly reproducible, so a verdict of \textbf{Results
Replicated} can confidently be given.

\section{Appendix 376 Evaluation
Report}\label{appendix-376-paper-558-evaluation-report}

\subsection{Artifact A1}\label{artifact-a1-9}

\subsubsection{Hardware and Software
Environment}\label{hardware-and-software-environment-18}

\begin{itemize}
\tightlist
\item
  \textbf{Location}: TACC
\item
  \textbf{Node type}: P100
\item
  \textbf{OS Image}: CC-Ubuntu22.04-CUDA (officially supported)
\item
  \textbf{CPUs}: 2
\item
  \textbf{Threads}: 48
\item
  \textbf{RAM Size}: 128 GiB
\item
  \textbf{GPU}: 2x Nvidia P100
\end{itemize}

\subsubsection{Reproduction Attempt}\label{reproduction-attempt-18}

\begin{itemize}
\tightlist
\item
  \textbf{Date}: 2024-08-09
\item
  \textbf{Time}: 10:03
\end{itemize}

\subsubsection{Steps performed}\label{steps-performed-18}

\begin{enumerate}
\def\labelenumi{\arabic{enumi}.}
\tightlist
\item
  Opened Trovi artifact, failed to open directly from the link provided
  in the paper (this is likely a chameleon problem). Workaround:
  download the artifact and upload it to a fresh Jupyter notebook.
\item
  Executed the notebook step by step. Everything worked nearly as
  expected, see notes below.
\end{enumerate}

Total reproduction time: 90 minutes

\subsubsection{Notes}\label{notes-18}

\begin{itemize}
\tightlist
\item
  The workbook has no instruction for manual reservation/lease. While
  the automatic one works fine, this is a potential issue for users who
  want to reserve a specific server or who already have a reservation.

  \begin{itemize}
  \tightlist
  \item
    Furthermore, configuration is not centralized and not explained
    well. This should be improved in the template.
  \end{itemize}
\item
  There are very few outputs from the provided scripts. Especially for
  long-running operations and their completion, it is vital for the user
  to know what is happening and \emph{that} something is happening.
\item
  There are some typos in the script and its output, e.g.~``benchamrk''
\end{itemize}

\subsubsection{Verdict}\label{verdict-18}

This is an extremely nice demonstration of a Trovi artifact. The
notebook is well-structured and easy to follow. The only issue is the
lack of information about the reservation system and the lack of output
from the scripts. This is a minor issue, but it could be improved.
Nevertheless, this artifact is a very clear
\texttt{Results\ Replicated}.

\section{Appendix 407 Evaluation
Report}\label{appendix-407-paper-512-evaluation-report}

\subsection{Artifact A1}\label{artifact-a1-10}

\subsubsection{Hardware and Software
Environment}\label{hardware-and-software-environment-19}

\begin{itemize}
\tightlist
\item
  \textbf{Location}: TACC
\item
  \textbf{Node type}: MI100
\item
  \textbf{OS Image}: CC-Ubuntu22.04-CUDA (officially supported)
\item
  \textbf{CPUs}: 2
\item
  \textbf{Threads}: 256
\item
  \textbf{RAM Size}: 256 GiB
\item
  \textbf{GPU}: 1x AMD MI100
\end{itemize}

\subsubsection{Reproduction Attempt}\label{reproduction-attempt-19}

\begin{itemize}
\tightlist
\item
  \textbf{Date}: 2024-08-14
\item
  \textbf{Time}: 13:31
\end{itemize}

\subsubsection{Steps performed}\label{steps-performed-19}

\begin{enumerate}
\def\labelenumi{\arabic{enumi}.}
\tightlist
\item
  Cloned repository using
  \texttt{git\ clone\ https://github.com/Nano-TCAD/QuaTrEx.git}
\item
  Ran \texttt{install\_general.sh} script, rebooted the system
\item
  Ran \texttt{install\_python.sh} script, added \texttt{conda} to the
  path
\item
  Ran \texttt{install\_rocm\_driver.sh} script, rebooted the system
\item
  Ran \texttt{install\_rocm\_software.sh} script
\item
  Ran \texttt{install\_ompi.sh} script
\item
  Ran \texttt{install\_cupy.sh} script
\item
  Ran \texttt{install\_mpi4py.sh} script
\item
  Installed python package locally using
  \texttt{cd\ ..;\ python3\ -m\ pip\ install\ -\/-editable\ .}
\end{enumerate}

\subsubsection{Notes}\label{notes-19}

\begin{itemize}
\tightlist
\item
  This could easily be automated using a Jupyter notebook on Chameleon.
  I like the separation of concerns with the install scripts, but
  manually reconnecting after reboot is tedious.
\end{itemize}

\subsubsection{Verdict}\label{verdict-19}

This is a strangely defined artifact, only containing the software setup
and no actual experiments. The setup is well documented and easy to
follow, but the lack of experiments makes it hard to evaluate the
artifact.

\subsection{Artifact A2}\label{artifact-a2-4}

\subsubsection{Hardware and Software
Environment}\label{hardware-and-software-environment-20}

\emph{Same as A1}

\subsubsection{Reproduction Attempt}\label{reproduction-attempt-20}

\emph{Same as A1}

\subsubsection{Steps performed}\label{steps-performed-20}

Downloaded \texttt{small.zip} and \texttt{large.zip} from the provided
Zenodo link. Unzipped both files to obtain \texttt{large} and
\texttt{small} folders.

\subsubsection{Notes}\label{notes-20}

\begin{itemize}
\tightlist
\item
  The guidelines really need to include something about artifact
  definition. An artifact should mirror some results from the paper, not
  just software setup.
\end{itemize}

\subsubsection{Verdict}\label{verdict-20}

Again, this is a strangely defined artifact, only containing the
software setup and no actual experiments, which this time is only
downloading of data. The lack of experiments makes it hard to evaluate
the artifact.

\subsection{Artifact A3}\label{artifact-a3-1}

\subsubsection{Hardware and Software
Environment}\label{hardware-and-software-environment-21}

\emph{Same as A1}

\subsubsection{Reproduction Attempt}\label{reproduction-attempt-21}

\emph{Same as A1}

\subsubsection{Steps performed}\label{steps-performed-21}

\begin{enumerate}
\def\labelenumi{\arabic{enumi}.}
\tightlist
\item
  Downloaded third artifact from zenodo link.
\item
  Unzipped the file to obtain \texttt{experiments} folder. Moved
  \texttt{small} and \texttt{large} into it according to the
  instructions.
\item
  Executed \texttt{experiments.sh} script, prompting that
  \texttt{matplotlib} is not installed.
\item
  Installed \texttt{matplotlib} using
  \texttt{python3\ -m\ pip\ install\ matplotlib}
\item
  Re-ran \texttt{experiments.sh} script. Unfortunately, the following
  error occurred:
  \texttt{AttributeError:\ \textquotesingle{}csc\_matrix\textquotesingle{}\ object\ has\ no\ attribute\ \textquotesingle{}H\textquotesingle{}}
\item
  After encountering the error, the QuaTrEx repository was re-downloaded
  from the Zenodo link and the package was reinstalled. The same error
  persisted. See the full trace below:
\end{enumerate}

{ \scriptsize
\begin{verbatim}
Traceback (most recent call last):
  File "/home/cc/apdx407/experiments/gw_sc24_test.py", line 96, in <module>
    hamiltonian_obj = OMENHamClass.Hamiltonian(hamiltonian_path,
                      no_orb, Vappl = Vappl, potential_type = 'atomic',
                      bias_point = 13, rank = rank, layer_matrix = '/Layer_Matrix.dat')
  File "/home/cc/apdx407/Nano-TCAD-QuaTrEx-830126c/quatrex/OMEN_structure_matrices
                                               /OMENHamClass.py", line 106, in __init__
    self.hermitean = self.check_hermitivity(tol=1e-6)
  File "/home/cc/apdx407/Nano-TCAD-QuaTrEx-830126c/quatrex/OMEN_structure_matrices
                                              /OMENHamClass.py", line 432, in check_hermitivity
    err44 = np.max(np.absolute(self.Hamiltonian['H_4'] - self.Hamiltonian['H_4'].H))
\end{verbatim}
}
\begin{enumerate}
\def\labelenumi{\arabic{enumi}.}
\setcounter{enumi}{6}
\tightlist
\item
  It turns out, the \texttt{.H} attribute denotes a Hermitian transpose,
  which is not available for \texttt{csc\_matrix} objects. The error was
  fixed by replacing
  \texttt{self.Hamiltonian{[}\textquotesingle{}H\_4\textquotesingle{}{]}.H}
  with
  \texttt{self.Hamiltonian{[}\textquotesingle{}H\_4\textquotesingle{}{]}.conj().T}
  in the \texttt{OMENHamClass.py} file. This is likely due to a version
  incompatibility between used packages.
\item
  After fixing the error, the script ran successfully. (Time taken: 5:30
  min). The results of subsequently running the \texttt{stats.py} script
  are as follows:
\end{enumerate}

{ \scriptsize
\begin{verbatim}
(base) cc@sor-apdx-eval-mi100-2:~/apdx407/experiments$ python3 stats.py small_e32_n01.csv
Kernel G-K1.1: 1.8536418584990315 seconds
Kernel G-K1.2: 2.0645459724983084 seconds
Kernel W-K1.1: 3.386408284510253 seconds
Kernel W-K1.2: 2.5257889770073234 seconds
Kernel iteration: 10.343545081501361 seconds
(base) cc@sor-apdx-eval-mi100-2:~/apdx407/experiments$ python3 stats.py large_e16_n01.csv
Kernel G-K1.1: 0.9469771994990879 seconds
Kernel G-K1.2: 1.5243427760069608 seconds
Kernel W-K1.1: 6.45517437250237 seconds
Kernel W-K1.2: 5.5386661130032735 seconds
\end{verbatim}
}

These results are very roughly consistent with the results from the
paper, with W-K1.1 taking the longest in both batch sizes. However,
W-K1.2 is slower than both G-K1 kernels, which is not consistent with
the paper. Repeating the experiment three times did not change the
results.

\subsubsection{Notes}\label{notes-21}

\begin{itemize}
\tightlist
\item
  While it is nice that the artifact contains the original data and
  plotting scripts, this could (and should) be included in the AE
  report.
\item
  I am not sure whether stating that the resuls should be qualitatively
  compared is a good idea. More explanation about the expected results
  would make this a more objective evaluation.
\end{itemize}

\subsubsection{Verdict}\label{verdict-21}

With some small issues, this artifact is mostly reproducible. The error
encountered was fixed by replacing the \texttt{.H} attribute with
\texttt{conj().T}. The results are mostly consistent with the paper, but
there are some discrepancies. The overall verdict is
\texttt{Results\ Replicated}, but with some minor issues.

\section{Appendix 428 Evaluation
Report}\label{appendix-428-paper-230-evaluation-report}

\subsection{Artifact A1}\label{artifact-a1-11}

\subsubsection{Hardware and Software
Environment}\label{hardware-and-software-environment-22}

\begin{itemize}
\tightlist
\item
  \textbf{Location}: TACC
\item
  \textbf{Node type}: icelake\_r650
\item
  \textbf{OS Image}: CC-Ubuntu22.04 (officially supported)
\item
  \textbf{CPUs}: 2
\item
  \textbf{Threads}: 160
\item
  \textbf{RAM Size}: 256 GiB
\end{itemize}

\subsubsection{Reproduction Attempt}\label{reproduction-attempt-22}

\begin{itemize}
\tightlist
\item
  \textbf{Date}: 2024-07-29
\item
  \textbf{Time}: 13:57
\end{itemize}

\subsubsection{Steps performed}\label{steps-performed-22}

\textbf{Note: this appendix is not part of the SoR project, but was
evaluated as part of the official artifact evaluation process, thus the
documentation here is not consistent with the other appendices since the
authors only requested an \texttt{Artifact\ Functional} badge.}

There is no artifact evaluation appendix for artifact A2 , and the
artifact evaluation appendix for A1 is too broad, since it includes no
specific instructions on performing the actual experiments. Thus, this
part of the evaluation report focuses on the overall setup of the
software. The installation procedure was not very self-contained,
requiring the use of external resources to install all necessary
software. For example, the spack installation on the used Ubuntu 22.04
image required the installation of external libraries and non-trivial
configuration that was neither outlined in the appendices nor in the
repository. Furthermore, the installation instructions from the appendix
deviate from those in the repository, with the former not mentioning a
specific commit/branch to use. Thus, for the evaluation, the repository
instructions for installation with spack were followed. We have also
managed a successful development build of the software (as outlined in
the appendix) on the latest commit of the main branch.

As the provisioning of the necessary hardware and software (partial
software downgrade is necessary for some experi- ments) exceeds the
scope of this evaluation, we will focus on the official criteria for the
artifact functional badge the authors applied for: -
\emph{Documentation: Are the artifacts sufficiently documentedto enable
them to be exercised by readers of the paper?} Both yes and no. While
the folder structure is well-documented, the installation process could
have been more detailed and self-contained, and would have benefited
from a more detailed, replicable workflow on a single machine. -
\emph{Completeness: Do the submitted artifacts include all the key
components described in the paper?}: Yes, the raw data from the
experiments is present in the artifact. - \emph{Exercisability: Do the
submitted artifacts include the scripts and data needed to run the
experiments described in the paper, and can the software be successfully
executed?} As stated, the provisioning of the hardware and software for
full replication exceeds the scope of this evaluation. However, while
with some difficulties that could be avoided with a more detailed
installation guide, the software was successfully built in both
development and release mode and the tests were executed. It remains
however questionable whether an interested reader could replicate the
experiments without additional help.

To conclude, while the artifact fulfills some criteria for the artifact
functional badge, of the things replicable in small- scale settings,
there were too many issues to warrant more than an \emph{artifact
available} badge.

\subsubsection{Notes}\label{notes-22}

\begin{itemize}
\tightlist
\item
  There is no artifact evaluation appendix for this paper, which is
  technically required for the artifact to be considered for any badge
  but \texttt{Artifact\ Available}.
\end{itemize}

\subsubsection{Verdict}\label{verdict-22}

The artifact is available, but the lack of a detailed installation guide
and the need for very specific, unrealistic hardware requirements
prevent it from receiving any badge other than
\texttt{Artifact\ Available}.

\section{Appendix 466 Evaluation
Report}\label{appendix-466-paper-544-evaluation-report}

\subsection{Artifacts A1-A5}\label{artifacts-a1-a5}

\subsubsection{Hardware and Software
Environment}\label{hardware-and-software-environment-23}

\begin{itemize}
\tightlist
\item
  \textbf{Location}: UC
\item
  \textbf{Node type}: V100
\item
  \textbf{OS Image}: CC-Ubuntu22.04-CUDA (officially supported)
\item
  \textbf{CPUs}: 2
\item
  \textbf{Threads}: 80
\item
  \textbf{RAM Size}: 128 GiB
\item
  \textbf{GPU}: NVIDIA V100
\end{itemize}

\subsubsection{Reproduction Attempt}\label{reproduction-attempt-23}

\begin{itemize}
\tightlist
\item
  \textbf{Date}: 2024-08-14
\item
  \textbf{Time}: 14:53
\end{itemize}

\subsubsection{Steps performed}\label{steps-performed-23}

*Note: the artifacts in this appendix are very granular, and the
software and hardware setup is similar for all of them. For this reason,
we describe the evaluation of all artifacts in this section.

\begin{enumerate}
\def\labelenumi{\arabic{enumi}.}
\tightlist
\item
  Installed \texttt{conda} using the instructions from the respective
  website.
\item
  Cloned the repository, built and activated the environment according
  to the AE instructions:
\end{enumerate}

\begin{Shaded}
\begin{Highlighting}[]
\FunctionTok{git}\NormalTok{ clone https://github.com/JimZeyuYang/GPU\_Power\_Benchmark.git}
\BuiltInTok{cd}\NormalTok{ GPU\_Power\_Benchmark}
\ExtensionTok{conda}\NormalTok{ env create }\AttributeTok{{-}p}\NormalTok{ ./env }\AttributeTok{{-}f}\NormalTok{ bin/project\_env.yml}
\BuiltInTok{source}\NormalTok{ activate ./env}
\end{Highlighting}
\end{Shaded}

\begin{enumerate}
\def\labelenumi{\arabic{enumi}.}
\setcounter{enumi}{2}
\tightlist
\item
  \textbf{Artifact A1}: Ran the experiment with the the parameters
  \texttt{./source/benchmark.py\ -b\ -p\ -e\ 1}

  \begin{itemize}
  \tightlist
  \item
    Comparing \texttt{power\_update\_freq.jpg} to Fig. 6 in the paper,
    the results are similar for the V100 GPU, considering that both are
    histograms and a lot of the measurements seem to fall between two
    windows, leading to a similar plot area in the 20ms power update
    period.
  \end{itemize}
\item
  \textbf{Artifact A2}: Using the previous run, we can evaluate the
  transient response of the GPU power consumption in the
  \texttt{result.jpg} plots for the respective load levels.

  \begin{itemize}
  \tightlist
  \item
    The results are similar to the first response (A100) in the paper as
    they exhibit a similar pattern.
  \end{itemize}
\item
  \textbf{Artifact A3}: \emph{Did not run because no PMD is present in
  the provided hardware}.
\item
  \textbf{Artifact A4}: Ran the experiment with the the parameters
  \texttt{./source/benchmark.py\ -b\ -p\ -e\ 2}

  \begin{itemize}
  \tightlist
  \item
    The results show a similar behavior to the A100 GPU in Figure 10 of
    the paper, with the \texttt{nvidia-smi} power measurements swinging
    up and down despite constant GPU load.
  \end{itemize}
\item
  \textbf{Artifact A5}: Ran the experiment with the the parameters
  \texttt{./source/benchmark.py\ -b\ -p\ -e\ 5}

  \begin{itemize}
  \tightlist
  \item
    Again, since there is no PMD present, there is no ground truth to
    compare the results to. However, the results exhibit similar trends
    to the results reported in Fig. 18 from the paper.
  \end{itemize}
\end{enumerate}

\subsubsection{Verdict}\label{verdict-23}

It is hard to judge the quality of the artifacts without a ground truth
to compare them to. However, the results are consistent with the paper
and the artifacts are well documented. Due to the nice setup and the
similarity of the trends observed, a verdict of
\texttt{Results\ Replicated} is warranted.

\section{Appendix 467 Evaluation
Report}\label{appendix-467-paper-547-evaluation-report}

\subsection{Artifact A1}\label{artifact-a1-12}

\subsubsection{Hardware and Software
Environment}\label{hardware-and-software-environment-24}

\begin{itemize}
\tightlist
\item
  \textbf{Location}: TACC
\item
  \textbf{Node type}: P100
\item
  \textbf{OS Image}: CC-Ubuntu22.04-CUDA (officially supported)
\item
  \textbf{CPUs}: 2
\item
  \textbf{Threads}: 48
\item
  \textbf{RAM Size}: 128 GiB
\end{itemize}

\subsubsection{Reproduction Attempt}\label{reproduction-attempt-24}

\begin{itemize}
\tightlist
\item
  \textbf{Date}: 2024-08-17
\item
  \textbf{Time}: 12:10
\end{itemize}

\subsubsection{Steps performed}\label{steps-performed-24}

\begin{enumerate}
\def\labelenumi{\arabic{enumi}.}
\tightlist
\item
  Connected to the server \emph{with port forwarding of port 8889} for
  local access to the Jupyter notebook.
\item
  Cloned the repository:
\end{enumerate}

\begin{Shaded}
\begin{Highlighting}[]
\ExtensionTok{$}\NormalTok{ git clone }\DataTypeTok{\textbackslash{}}
\NormalTok{https://github.com/IBM/LLM{-}performance{-}prediction.git}
\BuiltInTok{cd}\NormalTok{ LLM{-}performance{-}prediction}
\end{Highlighting}
\end{Shaded}

\begin{enumerate}
\def\labelenumi{\arabic{enumi}.}
\setcounter{enumi}{2}
\tightlist
\item
  As CUDA is already installed, we only need to install docker, for
  which the script \texttt{docker/install\_docker.sh} is provided, which
  is first changed to be executable using \texttt{chmod\ +x}.
\item
  Built the docker container using the provided Dockerfile with
  \texttt{docker\ build}
\item
  Ran the docker image with GPU support and port forwarding for Jupyter
  notebook access:
\end{enumerate}

\begin{Shaded}
\begin{Highlighting}[]
\FunctionTok{sudo}\NormalTok{ docker run }\AttributeTok{{-}it} \DataTypeTok{\textbackslash{}}
\NormalTok{{-}{-}gpus all }\DataTypeTok{\textbackslash{}}
\NormalTok{{-}v ./preprocess\_data:/app/preprocess\_data }\DataTypeTok{\textbackslash{}}
\NormalTok{{-}v ./predict\_performance:/app/predict\_performance }\DataTypeTok{\textbackslash{}}
\NormalTok{{-}p 8889:8889 }\DataTypeTok{\textbackslash{}}
\NormalTok{llm{-}pilot}
\end{Highlighting}
\end{Shaded}

\begin{enumerate}
\def\labelenumi{\arabic{enumi}.}
\setcounter{enumi}{5}
\tightlist
\item
  Executed the following steps in the \texttt{Preprocess\_data.ipynb}
  notebook:

  \begin{enumerate}
  \def\labelenumii{\arabic{enumii}.}
  \tightlist
  \item
    Imported necessary libraries (cell 1)
  \item
    Read and preprocessed the data (cell 2)

    \begin{itemize}
    \tightlist
    \item
      Warning Message:
      \texttt{Notebook\ preprocess\_data.ipynb\ is\ not\ trusted} What
      does this mean? Copilot:
    \end{itemize}
  \item
    Trusted the notebook and executed first two cells again.
  \item
    Defined \texttt{plot\_performance} function for plotting (cell 3)
  \item
    Plotted results (cell 4).
  \end{enumerate}
\item
  All figures match the expected results. While there are differences in
  the binary files, the figures are the same, except spacing around the
  axes.
\end{enumerate}

\subsubsection{Notes}\label{notes-23}

\begin{itemize}
\tightlist
\item
  Again, providing scripts for dependency install and explicit commands
  is good practice.
\item
  There are no outputs for the long running operation in cell 2 of
  preprocess data (6.2/6.3). This is a bit concerning as it is not clear
  if the operation is running or not.
\item
  \textbf{Guidelines}: Clear outputs of Jupyter notebook cells are
  important for reproducibility. If the notebook already contains all
  outputs, it is not immediately clear if the operation has already been
  executed or not.
\end{itemize}

\subsubsection{Verdict}\label{verdict-24}

This is a very nice example of using a Jupyter notebook for
reproducibility. The notebook is well documented and the results are as
expected. While there are some small issues, the overall verdict is a
clear \texttt{Results\ Replicated}.

\subsection{Artifact A2}\label{artifact-a2-5}

\subsubsection{Hardware and Software
Environment}\label{hardware-and-software-environment-25}

\begin{itemize}
\tightlist
\item
  \textbf{Location}: TACC
\item
  \textbf{Node type}: P100
\item
  \textbf{OS Image}: CC-Ubuntu22.04 CUDA (officially supported)
\item
  \textbf{CPUs}: 2
\item
  \textbf{Threads}: 48
\item
  \textbf{RAM Size}: 128 GiB
\end{itemize}

\subsubsection{Reproduction Attempt}\label{reproduction-attempt-25}

\begin{itemize}
\tightlist
\item
  \textbf{Date}: 2024-08-17
\item
  \textbf{Time}: 13:10
\end{itemize}

\subsubsection{Steps performed}\label{steps-performed-25}

\begin{enumerate}
\def\labelenumi{\arabic{enumi}.}
\tightlist
\item
  Using the setup from the previous artifact, connected to the Jupyter
  notebook \texttt{Predict\_LLM\_performance.ipynb}.
\item
  Executed the following steps in the notebook:

  \begin{enumerate}
  \def\labelenumii{\arabic{enumii}.}
  \tightlist
  \item
    Imported necessary libraries (cell 1)
  \item
    Loaded and encoded preprocessed data (cells 2-4):

    \begin{itemize}
    \tightlist
    \item
      Read and preprocessed the data (cell 2)
    \item
      Added sample weights (cell 3)
    \item
      Added reference performance evaluations (cell 4)
    \end{itemize}
  \item
    Defined constants and utilities for performance estimation (cell 5)
  \item
    Ran LLM-Pilot (cell 6), \textbf{1.2 minutes}
  \item
    Made predictions using reference methods:

    \begin{itemize}
    \tightlist
    \item
      Ran Random Forest (cell 7) \textbf{5.5 minutes}
    \item
      Ran PARIS (cell 8) \textbf{13.4 minutes}
    \item
      Ran Selecta (cells 9-10) \textbf{17.2 minutes}
    \item
      Ran PerfNET (cell 11) \textbf{14 minutes}
    \item
      Ran PerfNETV2 (cell 12) \textbf{21 minutes}
    \item
      Ran Morphling (cells 13-14) \textbf{7.7 minutes}
    \end{itemize}
  \item
    Analyze static policy baselines and select best one (cell 15)
  \item
    Make GPU recommendations based on LLM-Pilot and baseline results
  \item
    Plot results
  \end{enumerate}
\item
  While the results of LLM-Pilot are as expected, the results of some
  other methods (namely Perfnet, Morphling, and PerfNetV2) are not, but
  it is described in the AD/AE appendix that especially these
  implementations are sensitive to hardware changes. Other than that,
  the produced figure is identical to the expected one.
\end{enumerate}

\subsubsection{Notes}\label{notes-24}

\begin{itemize}
\tightlist
\item
  No outputs for the operations in cells 2-5
\item
  Nice output for the Prediction operations.
\end{itemize}

\subsubsection{Verdict}\label{verdict-25}

While some of the baseline results are not as expected, the results from
the paper have been replicated, and the notebooks and software setup are
well documented and executed. This artifact deserves a
\texttt{Results\ Replicated} badge.

\subsection{Overall Verdict}\label{overall-verdict-5}

Both artifacts have been successfully replicated. The notebooks are well
documented and the results are as expected. The only issue is the lack
of outputs for some operations, which can be improved for better
reproducibility. Overall, this is a clear \texttt{Results\ Replicated}
for the appendix.

\section{Appendix 482 Evaluation
Report}\label{appendix-482-paper-136-evaluation-report}

\subsection{Artifact A1}\label{artifact-a1-13}

\subsubsection{Hardware and Software
Environment}\label{hardware-and-software-environment-26}

\begin{itemize}
\tightlist
\item
  \textbf{Location}: TACC
\item
  \textbf{Node type}: icelake\_r650
\item
  \textbf{OS Image}: CC-Ubuntu22.04 (officially supported)
\item
  \textbf{CPUs}: 2
\item
  \textbf{Threads}: 160
\item
  \textbf{RAM Size}: 256 GiB
\end{itemize}

\subsubsection{Reproduction Attempt}\label{reproduction-attempt-26}

\begin{itemize}
\tightlist
\item
  \textbf{Date}: 2024-08-12
\item
  \textbf{Time}: 14:00
\end{itemize}

\subsubsection{Steps performed}\label{steps-performed-26}

\begin{enumerate}
\def\labelenumi{\arabic{enumi}.}
\tightlist
\item
  Cloned repository from
  \url{https://anonymous.4open.science/r/coaxial\_artifact-SC/}
\item
  Proceeded according to the README instructions:

  \begin{itemize}
  \tightlist
  \item
    \texttt{cd\ TRACES}
  \item
    created a \texttt{traces} folder and ran
    \texttt{perl\ download\_traces.pl\ -\/-csv\ artifact\_traces.csv\ -\/-folder\ traces}
  \end{itemize}
\item
  Trace downloading completed after more than 10 hours (overnight), the
  successful download of only 138/233 is reported.
\end{enumerate}

\begin{verbatim}
================================
Trace downloading completed
Downloaded 138/233 traces
================================
\end{verbatim}

\textbf{Trace downloading failed, this makes the reproduction of the
results impossible. We will continue to describe the process of building
the tools, but running without traces is not possible.} 4. Built
DRAMSim3 using \texttt{make} in the respective directory 5. Building
ChampSim3 using \texttt{make} failed. The mentioned \texttt{config.sh}
script is not available in the artifact, and \texttt{make\ clean}
removes header files required for building the tool. Using the provided
\texttt{config.py} script, a new Makefile was generated, which
successfully built the tool. 6. Executing ChampSim3 requires setting the
environment variable \texttt{LD\_PRELOAD} (thanks to the reviewers for
pointing this out!), facilitating the execution of the tool. With the
review hardware, this simulation did not terminate within 12 hours.

\subsubsection{Notes}\label{notes-25}

\begin{itemize}
\tightlist
\item
  The artifact description is very incomplete, and the anonymized GitHub
  is not very navigable. Also, folder names are written in caps,
  different than the paths described in the README.
\item
  The network connection of the trace providing servers is extremely
  slow, seldomly exceeding 500 KB/s. This makes the download of the
  traces very slow. It is quite audatious to state that the download
  time is dependent on the network connection of the user.
\item
  Particularly referring to the README in several steps of the
  reproduction process is not very user-friendly. The artifact
  evaluation appendix should be (mostly) self-contained.
\end{itemize}

\subsubsection{Verdict}\label{verdict-26}

Sadly, the reproduction of the results was not possible due to the
failed download of the traces. The artifact description is incomplete,
and the network connection of the trace providing servers is extremely
slow. Moreover, building the tools is not straightforward, not
well-documented, and the executable did not terminate on our hardware
within 12 hours. The artifact should be improved to make the
reproduction of the results possible. This only warrants a
\texttt{Artifact\ Available} badge.

\section{Appendix 483 Evaluation
Report}\label{appendix-483-paper-434-evaluation-report}

\subsection{Artifact A1}\label{artifact-a1-14}

\subsubsection{Hardware and Software
Environment}\label{hardware-and-software-environment-27}

\begin{itemize}
\tightlist
\item
  \textbf{Location}: TACC
\item
  \textbf{Node type}: icelake\_r650
\item
  \textbf{OS Image}: CC-Ubuntu22.04 (officially supported)
\item
  \textbf{CPUs}: 2
\item
  \textbf{Threads}: 160
\item
  \textbf{RAM Size}: 256 GiB
\end{itemize}

\subsubsection{Reproduction Attempt}\label{reproduction-attempt-27}

\begin{itemize}
\tightlist
\item
  \textbf{Date}: 2024-08-19
\item
  \textbf{Time}: 20:37
\end{itemize}

\subsubsection{Steps performed}\label{steps-performed-27}

\textbf{Note: Due to the lack of access to Intel Tofino SDK, the
evaluation of most parts of the artifact was not possible.} We will thus
focus on all the steps that could be performed without the SDK.

\begin{enumerate}
\def\labelenumi{\arabic{enumi}.}
\tightlist
\item
  Installed VirtualBox 6.1.50 using the \texttt{apt} package manager.
\item
  Downloaded the VirtualBox image from the provided Zenodo link.
\item
  Imported the image into VirtualBox using \texttt{VBoxManage\ import}.
\item
  Configured the machine using
  \texttt{VBoxManage\ modifyvm\ netcl-ubuntu-22.04\ -\/-natpf1\ \textquotesingle{}guestssh,tcp,,2222,,22"},
  forwarding port \texttt{22} of the guest to port \texttt{2222} of the
  host.
\item
  SSH-ed into the guest machine using
  \texttt{ssh\ -p\ 2222\ nc@localhost}
\item
  Executed \texttt{\textasciitilde{}/netcl\_paper\_artifact/run\_all.sh}
  to execute all artifacts. The script prompted that Tofino SDK and
  consequently the P4 compiler is not installed, and skipped the
  respective artifacts. The results of the computation are shown below:
\end{enumerate}

\textbf{Table 3}*

{ \footnotesize
\begin{verbatim}
+-------------- + ---- + ------ + ---------- + ------ + ------ + ----------- + ------ +
| program        | ncl  | p4     | reduction  | ncl.i  | p4.i   | reduction.i | p4-gen |
+ -------------- + ---- + ------ + ---------- + ------ + ------ + ----------- + ------ +
| agg            | 38   | 1360   | 35.79      | 33     | 1283   | 33.76       | 1278   |
| calculator     | 25   | 139    | 5.56       | 25     | 139    | 5.56        | 512    |
| netcache       | 97   | 692    | 7.13       | 88     | 1518   | 15.65       | 1592   |
| paxos-acceptor | 38   | 230    | 6.05       | 34     | 208    | 5.47        | 614    |
| paxos-leader   | 26   | 214    | 8.23       | 23     | 192    | 7.38        | 522    |
| paxos-learner  | 35   | 241    | 6.89       | 33     | 219    | 6.26        | 637    |
| paxos-one      | 76   | 387    | 5.09       | 70     | 365    | 4.80        | -      |
+ -------------- + ---- + ------ + ---------- + ------ + ------ + ----------- + ------ +
\end{verbatim}
}

The reduction results of Table 3 are not exactly the same as the ones in
the paper, but they are very close. The results are considered to be
reproduced.

\textbf{Table 4}

{ \footnotesize
\begin{verbatim}
+ --- + ----- + ---------- + -------- + -------------- + ------------ + ------------- +
|     | agg   | calculator | netcache | paxos-acceptor | paxos-leader | paxos-learner |
+ --- + ----- + ---------- + -------- + -------------- + ------------ + ------------- +
| ncc | 0.72  | 0.69       | 0.78     | 0.71           | 0.69         | 0.70          |
| p4c |   --  |   --       |   --     |   --           |   --         |   --          |
| tot | 0.72  | 0.69       | 0.78     | 0.71           | 0.69         | 0.70          |
+ --- + ----- + ---------- + -------- + -------------- + ------------ + ------------- +
\end{verbatim}
}

The compilation times roughly match the ones in the paper, and are
considered to be reproduced.

\textbf{Table 6}

{ \tiny
\begin{verbatim}
+ ---------- + ---------- + ---------- + ---------- + -------------- + ------------ + ------------- +
|            | agg        | calculator | netcache   | paxos-acceptor | paxos-leader | paxos-learner |
+ ---------- + ---------- + ---------- + ---------- + -------------- + ------------ + ------------- +
| IR Allocas | 1 / 16b    | 1 / 8b     | 3 / 176b   | 0              | 1 / 8b       | 1 / 8b        |
| P4 Locvars | 4 / 80b    | 2 / 40b    | 17 / 349b  | 3 / 64b        | 2 / 40b      | 7 / 80b       |
+ ---------- + ---------- + ---------- + ---------- + -------------- + ------------ + ------------- +
\end{verbatim}
}

Several of the values in Table 6 are not the same as the ones in the
paper, but they are very close. Namely, the following do not match: - IR
Allocas \texttt{agg} (1 / 16b vs 0) - IR Allocas \texttt{netcache} (3 /
176 vs 3 / 88b) - P4 Locvars \texttt{netcache} (17 / 349b vs 17 / 466b)
- P4 Locvars \texttt{paxos-leader} (2 / 40b vs 1 / 8b) - P4 Locvars
\texttt{paxos-leader} (7 / 80b vs 9 / 96b)

\textbf{Figure 12} Figure 12 looks very similar to the one in the paper,
and is definitely considered to be reproduced.

\subsubsection{Notes}\label{notes-26}

\begin{itemize}
\tightlist
\item
  VirtualBox is \emph{incredibly tedious} to use. The authors definitely
  could have included a more detailed guide on how to use the software,
  as it is quite uncommon on Linux systems (especially headless ones).
\item
  The \texttt{run\_all.sh} is a well-written script that produces
  followable output and a nice visual representation of the results. It
  is a pity that it is so tedious to set up the environment to run it.
\end{itemize}

\subsubsection{Verdict}\label{verdict-27}

The results of the artifact are considered to be \textbf{reproduced}.
The lack of access to the Tofino SDK prevented the evaluation of some
parts of the artifact, but the results that could be evaluated were very
close to the ones in the paper. Also, once VirtualBox was set up, the
artifact was easy to run and produced the expected results in a clear
and concise manner.

\section{Appendix 496 Evaluation
Report}\label{appendix-496-paper-210-evaluation-report}

\subsection{Artifact A1}\label{artifact-a1-15}

\subsubsection{Hardware and Software
Environment}\label{hardware-and-software-environment-28}

\begin{itemize}
\tightlist
\item
  \textbf{Location}: TACC
\item
  \textbf{Node type}: icelake\_r650
\item
  \textbf{OS Image}: CC-Ubuntu22.04 (officially supported)
\item
  \textbf{CPUs}: 2
\item
  \textbf{Threads}: 160
\item
  \textbf{RAM Size}: 256 GiB
\end{itemize}

\subsubsection{Reproduction Attempt}\label{reproduction-attempt-28}

\begin{itemize}
\tightlist
\item
  \textbf{Date}: 2024-08-11
\item
  \textbf{Time}: 14:14
\end{itemize}

\subsubsection{Steps performed}\label{steps-performed-28}

\emph{No steps to be performed for this artifact, as the data is already
provided}

\subsubsection{Notes}\label{notes-27}

\begin{itemize}
\tightlist
\item
  I am not sure whether the authors made a good point in that the
  artifact is not supposed to be replicated.
\end{itemize}

\subsubsection{Verdict}\label{verdict-28}

I am not sure whether one can give a verdict for this artifact. The data
is provided, and the authors claim that the artifact is not supposed to
be replicated since it consists of a correctness check only. The authors
could at least have given more details about the execution in the AE
appendix.

\subsubsection{Hardware and Software
Environment}\label{hardware-and-software-environment-29}

\textbf{Same as A1}

\subsubsection{Reproduction Attempt}\label{reproduction-attempt-29}

\textbf{Same as A1}

\subsubsection{Steps performed}\label{steps-performed-29}

\begin{enumerate}
\def\labelenumi{\arabic{enumi}.}
\tightlist
\item
  Downloaded artifacts from zenodo under the given DOI
  (\url{https://zenodo.org/api/records/12594486/files-archive})
\item
  Executed the python scripts in the given order, each script ran
  successfully and wrote a CSV file to the \texttt{csv\_data} directory.

  \begin{itemize}
  \tightlist
  \item
    \texttt{python3\ our\_lattice\_surgery.py}
  \item
    \texttt{python3\ our\_heavy-hex.py}
  \item
    \texttt{python3\ our\_sycamore.py}
  \end{itemize}
\item
  The \texttt{sabre/sabre\_qft.py} script was executed with three
  different inputs. All scripts ran successfully.

  \begin{itemize}
  \tightlist
  \item
    \texttt{python3\ sabre\_qft.py\ ’N*N’} (\textasciitilde1:30 minutes)
  \item
    \texttt{python3\ sabre\_qft.py\ ’sycamore’} (\textasciitilde30
    seconds)
  \item
    \texttt{python3\ sabre\_qft.py\ ’heavy-hex’} (\textasciitilde1
    minute)
  \end{itemize}
\item
  The \texttt{draw\_figures.py} script requires the \texttt{pandas} and
  \texttt{matplotlib} packages, which are not stated as dependency.
  After installing the packages, the script runs as expected.
\item
  On a non-X11 machine, the script does not display the figures due to
  the use of \texttt{plt.show()}. Thus, the script was modified to save
  the figures as PNG files.
\item
  Comparing the resulting figures and table with the provided ones, the
  results are consistent. However, there are slight discrepancies in the
  figures, but the overall trend of the results is captured.
\end{enumerate}

\subsubsection{Notes}\label{notes-28}

\begin{itemize}
\tightlist
\item
  The authors know the replication is often executed on non-X11
  machines, so it would be helpful to save the figures as PNG files by
  default. This should be included in the guidelines.
\end{itemize}

\subsubsection{Verdict}\label{verdict-29}

The artifact was successfully reproduced. The results are consistent
with the provided ones, with slight discrepancies in the figures. The
scripts ran successfully, but some dependencies were not stated in the
documentation. This should nevertheless be considered a successful
reproduction.

\subsection{Overall Verdict}\label{overall-verdict-6}

If A1 is sufficient for being considered as a successful reproduction,
the overall verdict is positive.

\section{Appendix 506 Evaluation
Report}\label{appendix-506-paper-420-evaluation-report}

\subsection{Artifact A1}\label{artifact-a1-16}

\subsubsection{Hardware and Software
Environment}\label{hardware-and-software-environment-30}

\begin{itemize}
\tightlist
\item
  \textbf{Location}: TACC
\item
  \textbf{Node type}: icelake\_r650
\item
  \textbf{OS Image}: CC-Ubuntu22.04 (officially supported)
\item
  \textbf{CPUs}: 2
\item
  \textbf{Threads}: 160
\item
  \textbf{RAM Size}: 256 GiB
\end{itemize}

\subsubsection{Reproduction Attempt}\label{reproduction-attempt-30}

\begin{itemize}
\tightlist
\item
  \textbf{Date}: 2024-08-11
\item
  \textbf{Time}: 14:57
\end{itemize}

\subsubsection{Steps performed}\label{steps-performed-30}

\begin{enumerate}
\def\labelenumi{\arabic{enumi}.}
\tightlist
\item
  Downloaded Source from the GitHub repository
\item
  Executed \texttt{run\_baseline.py} \emph{without} Graphite
  \texttt{python3\ run\_baseline.py\ 0}
\item
  Executed \texttt{run\_parallel.py} \emph{without} Graphite
  \texttt{python3\ run\_parallel.py\ 0}
\item
  Executed \texttt{python3\ graphine\_discretized\_compilation\_par.py}
\item
  The described python file
  \texttt{neutral-atom-compilation/\\neutralatomcompilation/eldi\_generate\_data.py}
  is misconfigured:

  \begin{itemize}
  \tightlist
  \item
    Firstly, the package \texttt{neutralatomcompilation} it is located
    in is not listed as dependency, thus there are import errors.
  \item
    Secondly, even after installing \texttt{neturalatomcompilation}, the
    file \texttt{eldi\_generate\_data.py} does not run in a virtual
    environment due to invalid cross-import syntax error in the package.
  \end{itemize}
\item
  The docker container solves this issue, but running
  \texttt{python3\ setup.py\ install} is still required in the
  container. Then, the script runs successfully, but this requires
  access to an ARM64 machine.
\item
  However, in the docker container \texttt{run\_baseline.py} does not
  run because of a missing dependency (\texttt{bqskit}) that cannot be
  installed because of dependency conflicts.
\item
  Using partially the results from the container and partially from the
  host machine, the results are indeed reproducible. The software setup
  however had major issues, that should have been clarified by the
  authors.
\end{enumerate}

\subsubsection{Notes}\label{notes-29}

\begin{itemize}
\tightlist
\item
  The docker image does \emph{not} work on AMD64/x64! I'm getting the
  following error:
\end{itemize}

\begin{verbatim}
WARNING: The requested image's platform (linux/arm64) does not match
the detected host platform (linux/amd64/v4) and no specific platform
was requested
exec /bin/bash: exec format error
\end{verbatim}

Apparently, the image requires an ARM64 architecture, which is very
uncommon in server hardware and consequently does not work on AMD64. - I
don't see a reason for this project being a python package, as the
package is misconfigured and has no real use as a standalone package. -
\texttt{run\_baseline.py} and \texttt{run\_parallel.py} are not
reproducible in the docker container, even on an ARM machine since the
\texttt{bqskit} python package dependency is not installable due to
dependency conflicts.

\subsubsection{Verdict}\label{verdict-30}

Due to the numerous issues encountered, this is not a clear candidate
for \emph{Results Replicated}. The software setup is not reproducible
and the docker container is not usable on common server hardware. The
software setup should be clarified, and it is very unclear how the
authors intend to provide the \texttt{neutralatomcompilation} package.

\end{document}